\documentstyle[preprint,aps]{revtex}
\tighten
\draft
\widetext
\preprint{HUTP-98/A087, NUB 3190}
\bigskip
\bigskip
\begin{document}
\title{TeV-scale Supersymmetric Standard Model and Brane World}
\medskip
\author{Zurab Kakushadze\footnote{E-mail: 
zurab@string.harvard.edu}}
\bigskip\bigskip
\address{
Lyman Laboratory of Physics, Harvard University, Cambridge, MA 02138\\
and\\
Department of Physics, Northeastern University, Boston, MA 02115}
\date{December 17, 1998}
\bigskip
\medskip
\maketitle

\begin{abstract}
{}Recently we proposed a TeV-scale Supersymmetric Standard Model in which the
gauge coupling unification is as precise (at one loop) as in the MSSM, and occurs in the TeV range. One of the key ingredients of this model is the presence of new states neutral under $SU(3)_c\otimes SU(2)_w$ but charged under $U(1)_Y$ whose mass scale is around that of the electroweak Higgs doublets. In this paper we show that introduction of these states allows to gauge novel anomaly free discrete (as well as continuous) symmetries (similar to ``lepton'' and ``baryon'' numbers)  which suppress dangerous higher dimensional operators and stabilize proton. Moreover, we argue that these gauge symmetries are essential for successfully generating small neutrino masses via a recently proposed higher dimensional mechanism. Furthermore, the mass hierarchy between the up and down quarks ({\em e.g.}, $t$ {\em vs.} $b$) can be explained without appealing to large $\tan\beta$, and the $\mu$-term for the electroweak Higgs doublets (as well as for the new states) can be generated. We also discuss various phenomenological implications of our model which lead to predictions testable in the present or near future collider experiments. In particular, we point out that signatures of scenarios with high {\em vs.} low unification (string) scale might be rather different. This suggest a possibility that the collider experiments may distinguish between these scenarios even {\em without} a direct production of heavy Kaluza-Klein or string states.       
\end{abstract}
\pacs{}

\section{Introduction}

{}The discovery of D-branes \cite{polchi} has had a deep impact on our understanding of string theory - the only known theory that consistently incorporates quantum gravity. This may have important phenomenological implications. Thus, the Standard Model gauge fields (as well as the corresponding charged matter) may reside inside of $p\leq 9$ spatial dimensional $p$-branes (or a set of overlapping branes), while gravity lives in a larger (10 or 11) dimensional bulk of space-time. {\em A priori} this ``Brane World'' picture\footnote{For recent developments, see, {\em e.g.}, \cite{witt,lyk,TeV,dien,3gen,anto,ST,3gen1,TeVphen,BW}.} is a viable scenario, and in \cite{BW} it was actually argued to be a likely description of nature. In particular, combining together the requirements of gauge and gravitational coupling unification, dilaton stabilization and weakness of the Standard Model gauge couplings seems to suggest the brane world scenario (with the Standard Model fields living on branes with $3<p<9$) as a coherent picture for describing our universe \cite{BW}\footnote{The brane world picture in the effective field theory context was discussed in \cite{early,shif}.}. This is largely due to a much higher degree of flexibility of the brane world scenario compared with, say, the old perturbative heterotic framework.

{}Thus, for instance, in string theory 
the gauge and gravitational couplings are expected to unify 
(up to an order one factor due to various thresholds \cite{kap,BF}) at the string scale
$M_s=1/\sqrt{\alpha^\prime}$. In the brane world scenario {\em a priori} the string scale can be anywhere between the electroweak 
scale $M_{ew}$ and the Planck scale $M_P=1/\sqrt{G_N}$ (where $G_N$ is
the Newton's constant). If we assume that the bulk is ten dimensional,  then the four dimensional gauge and gravitational couplings 
scale as\footnote{For illustrative purposes here we are using the corresponding tree-level relations in Type I (or Type I$^\prime$) theory.} $\alpha\sim g_s/V_{p-3} M_s^{p-3}$ respectively $G_N\sim g_s^2/V_{p-3} V_{9-p} M_s^8$, where $g_s$ is the string coupling, and $V_{p-3}$ and $V_{9-p}$ are the compactification volumes inside and transverse to the $p$-branes, respectively. For $3<p<9$ there are two {\em a priori} independent volume factors, and, for the fixed gauge coupling $\alpha$ (at the
unification, that is, string scale) and four dimensional Planck scale $M_P$, the string scale is
not determined. This observation was used in \cite{witt} to argue that the gauge and gravitational coupling unification problem\footnote{For a review of the gauge and gravitational coupling unification problem in the perturbative heterotic string context, see, {\em e.g.}, \cite{Dienes}, and references therein. In the Type I context the discussions on this issue can be found in \cite{CKM,BW}.} can be ameliorated in this context by lowering the string scale $M_s$ down to the GUT scale $M_{\small{GUT}}\approx 2\times 10^{16}~{\mbox{GeV}}$ \cite{gut}\footnote{By the GUT scale here we mean the usual scale of gauge coupling unification in the MSSM obtained by extrapolating the LEP data in the assumption of the standard ``desert'' scenario.}. In \cite{lyk} it was noticed that $M_s$ can be further lowered all the way down to TeV.

{}In fact, in the brane world picture {\em a priori} the string scale can be as low as desired as long as it does not directly contradict current experimental data. In \cite{TeV} it was proposed that $M_s$ {\em as well as}\footnote{Note that the string scale $M_s$ cannot be too much lower than the fundamental Planck scale or else the string coupling $g_s$ as well as all the gauge couplings would come out too small contradicting the experimental data.} the fundamental (10 or 11 dimensional) Planck scale can be around TeV. The observed weakness of the four dimensional gravitational coupling then requires presence of at least two large ($\gg 1/M_s$) compact directions (which are transverse to the branes on which the Standard Model fields are localized). A general discussion of possible brane world embeddings
of such a scenario was given in \cite{anto,ST,BW}. 
In \cite{TeVphen} various non-trivial phenomenological 
issues were discussed in the context of the TeV-scale brane world scenario, and it was argued that this possibility does not appear to be automatically ruled out\footnote{For other recent 
works on TeV-scale string/gravity
scenarios, see, {\em e.g.}, \cite{dien,flavor1,flavor,neutrino,neutrino1,zura,related,AB}. For other scenarios with
lowered string scale and related works, see, {\em e.g.},
\cite{other}. TeV-scale compactifications were studied in \cite{quiros} in
the context of supersymmetry breaking.}. 

{}However, in such a scenario, as well as in any scenario where $M_s\ll
M_{\small{GUT}}$, the gauge coupling unification at $M_s$ would have to arise in a way
drastically different from the usual MSSM unification which occurs with a remarkable precision
\cite{gut}. 
It is then desirable to have a mechanism in the TeV-scale brane world scenario for lowering the unification scale. Moreover, it would also be necessary to find a concrete extension of the MSSM (where this new mechanism is realized) such that the unification prediction is just as precise as in the MSSM (at least at one loop). In fact, one could also require that such an extension {\em explain} why
couplings unify in the MSSM at all, that is, why the unification in the MSSM is {\em not} just an ``accident'' assuming that the TeV scale brane world scenario has the pretense of 
replacing the old framework.
 
{}In the brane world picture there
appears to exist a mechanism \cite{dien} for lowering the unification scale. Thus, let the ``size'' $R$ of the compact dimensions inside of the $p$-brane (where $p>3$) be somewhat large compared with $1/M_s$. Then the evolution of the gauge couplings above the Kaluza-Klein (KK) 
threshold $1/R$ is no longer logarithmic but power-like \cite{TV}. (This can alternatively be thought of as considering a higher dimensional theory at energy scales larger than $1/R$.) This observation was used in \cite{dien} to argue that the gauge coupling unification might occur at a scale (which in the brane world context would be identified with the string scale) much lower than $M_{\small{GUT}}$. 

{}In \cite{zura} we proposed a TeV-scale Supersymmetric Standard Model, which we would like to abbreviate as ``TSSM''. 
The gauge coupling unification in the TSSM indeed occurs via such a higher dimensional mechanism. Moreover, the unification in the TSSM is as precise (at one loop) as in the MSSM, and occurs in the TeV range\footnote{By the TeV range we do {\em not} necessarily mean that $M_s\sim 1~{\mbox{TeV}}$. In fact, as was argued in \cite{zura}, the gauge coupling unification constraints seem to imply that $M_s$ cannot really be lower than $10-100~{\mbox{TeV}}$.}. The TSSM also explains why the unification in the MSSM is not an accident - if the TSSM is indeed (a part of) the correct description of nature above the electroweak scale, then the gauge coupling unification in the MSSM is explained by the current lack of data which leads to the standard ``desert'' assumption. Moreover, as was pointed out in \cite{zura}, after a rather systematic search the TSSM is the only (simple) solution we found for the constraints guaranteeing that the gauge couplings unify as precisely (at one loop) as in the MSSM.    

{}It is therefore reasonable to take the TSSM as the starting point for addressing 
many other open questions that the TeV-scale brane world scenario faces. In this paper we
will focus on two of such issues: proton stability and neutrino masses. In fact, we will argue that the two issues are not completely unrelated, and provide rather tight constraints for model building. This appears to imply that the number of viable possibilities is rather limited which allows to explore them in a systematic fashion.     

{}Proton decay is one of the most obvious worries with the TeV-scale brane world scenario. Indeed, already in the context of the MSSM with the string scale $M_s$ around $M_{\small{GUT}}$ the proton decay problem becomes non-trivial: in the MSSM there {\em a priori} exist dimension 4 and 5 operators violating baryon as well as lepton numbers, and unless (at least some of) these operators are absent (or further suppressed), proton will decay with an unacceptable rate \cite{wein}. Operators with dimension 6 (and higher) are essentially safe in this context as they are suppressed by a factor of $1/M_s^2\sim 1/M_{\small{GUT}}^2$ (whereas the dimension 4 operators are not suppressed by any large scale, and the dimension 5 operators are suppressed only by a factor of $1/M_s\sim 1/M_{\small{GUT}}$ which is insufficient). However, in the TeV-scale brane world scenario the higher $d>4$ dimensional operators are suppressed only by a factor of $1/M_s^{d-4}$ where $M_s$ is around TeV, so that even dimension 6 operators as well as operators with even higher dimensions become dangerous. Then generically to ensure proton stability one must invoke some symmetries forbidding the dangerous operators at least up to some (rather high) dimension. Global continuous symmetries (such as baryon and lepton number conservation) or non-gauge discrete symmetries are not expected to work in this context for the following reason. First, generically quantum gravity effects (wormholes, {\em etc.}) are expected to violate such global symmetries and induce rapid proton decay \cite{gil}. Second, it is believed that there are no global symmetries in string theory \cite{global}\footnote{These two observations may not be completely unrelated.}. This then implies that we should consider either continuous or discrete {\em gauge} symmetries. 
Discrete gauge symmetries are believed to be stable under quantum gravity effects \cite{Krauss} (also see, {\em e.g.}, \cite{Wilczek}), and they do arise in string theory. Continuous gauge symmetries would protect proton from decaying much more ``efficiently'' than discrete gauge symmetries: thus, for instance, if we could gauge the baryon number, proton would be completely stable. However, continuous gauge symmetries have the disadvantage that they have to be broken at some scale (or else the corresponding massless gauge boson(s) would give rise to experimentally excluded long range forces) in such a way that we do not get the proton decay problem back. Discrete gauge symmetries are more advantageous from this viewpoint as there are no massless gauge bosons associated with them, and they can be exact. However, generically discrete gauge symmetries are expected to forbid dangerous operators only up to some finite dimension, so one has to explicitly check in a given situation whether the induced suppression is sufficient.

{}Proton stability in the TeV-scale brane world scenario has been briefly
addressed in \cite{dien,anto,ST,TeVphen}, and some possible ways out have
been suggested. Thus, in \cite{anto} two possible mechanism were
discussed. First, one can attempt to gauge the baryon number (which would
then somehow have to be broken on a ``distant'' brane). However, if we
gauge the baryon number with just the three generations of the Standard
Model, the corresponding $U(1)_B$ will be anomalous (in particular, the
mixed ${\mbox{Tr}}(SU(2)_w^2 U(1)_B)$ gauge anomaly does not cancel). One
can attempt to remedy this by adding the fourth generation with the baryon
charge assignments of opposite sign and 3 times the absolute value of those
for the usual generations \cite{anto}. This way one can cancel all the
anomalies\footnote{This $U(1)_B$ charge assignment can be viewed as
descending from a ``vector-like'' $SU(4)$ ``flavor'' symmetry with the
left-handed quarks transforming in ${\bf 4}$ of $SU(4)$, and the
left-handed anti-quarks transforming in ${\overline {\bf 4}}$. This $SU(4)$
``flavor'' gauge symmetry is anomaly free, which implies that the
corresponding $U(1)_B$ is also anomaly free.}. 
At any rate, {\em a priori} the precise mechanism of breaking $U(1)_B$ on a ``distant'' brane is not completely understood in the brane world context, so at present it is unclear whether this scenario can be considered conclusive. The possibility discussed in \cite{TeVphen} is essentially along the same lines: one can add three new generations with baryon number assignments
opposite to those of the first three generations, and assume that all of the new generations are ``top-like'' (that is, they become very heavy upon the electroweak symmetry breaking). 
The second possibility discussed in \cite{anto} is to gauge a discrete symmetry, in particular, the ${\bf Z}_3$ ``Generalized Baryon Parity'' symmetry introduced in \cite{IR}, which was argued in \cite{anto} to suppress proton decay to an acceptable rate. However, {\em a priori} it is unclear whether this discrete gauge symmetry is anomaly free. In fact, as we will point out in the following, it has ${\mbox{Tr}}({\bf Z}_3 U(1)^2_Y)$ and ${\mbox{Tr}}({\bf Z}^2_3 U(1)_Y)$ anomalies if we consider just the MSSM
spectrum augmented with chiral superfields corresponding to right-handed neutrinos. On the other hand, in the TSSM (which contains additional fields) this discrete symmetry can indeed be gauged consistently, {\em i.e.}, all the anomalies can be cancelled. (However, as we point out 
in the following, this symmetry alone does not forbid certain dimension 5 operators which would generate unacceptably large neutrino masses - see below.) 
Another possibility was discussed in \cite{ST}: a non-anomalous ``custodial'' $U(1)$ gauge symmetry (which can generically be present in Type I (Type I$^\prime$) compactifications) would protect proton from rapid decay. However, the coupling constant of this $U(1)$ is generically of the same order of magnitude as that of $U(1)_{em}$, so it would have to be broken at a relatively high scale giving us back the proton stability problem. In fact, such custodial $U(1)$'s generically might pose some difficulties in certain Type I (Type I$^\prime$) compactifications (such as the example considered in \cite{ST}) where the only fields charged under such a $U(1)$ are also charged under $SU(3)_c$: the custodial $U(1)$ cannot be broken unless $SU(3)_c$ is broken (and the scale at which the custodial $U(1)$ would have to be broken to be unobservable is generically too high to allow $SU(3)_c$ to be broken at such a scale). Finally, in \cite{dien} it was pointed out that a certain ${\bf Z}_2$ discrete symmetry forbids some of the operators leading to proton decay. However, this discrete symmetry does not forbid all such operators, in particular, those including derivatives, so just this ${\bf Z}_2$ symmetry is insufficient to solve the proton decay problem\footnote{This observation was also made by Nima Arkani-Hamed, Savas Dimopoulos and Gia Dvali.}. To summarize, it seems fair to say that the proton decay problem is an open question in the TeV-scale brane world scenario.         

{}In this paper we will show that in the TSSM one can gauge anomaly free
discrete (as well as continuous) symmetries which suppress dangerous higher
dimensional operators and stabilize proton. Our main emphasis will be on
discrete gauge symmetries which do not require any additional assumptions
(such as breaking a continuous gauge symmetry on a ``distant'' brane). In
particular, we will explicitly discuss such discrete gauge symmetries (more
concretely, a ${\bf Z}_3\otimes {\bf Z}_3$ discrete gauge symmetry) which forbid all dangerous operators possibly responsible for proton decay.
This ameliorates the proton stability problem in the TSSM in a {\em conclusive} way. Here we would like to stress that one of the key ingredients of the TSSM is the presence of new states neutral under $SU(3)_c\otimes SU(2)_w$ but charged under $U(1)_Y$ whose mass scale is around that of the electroweak Higgs doublets. These states are crucial for the gauge coupling unification in the TSSM along the lines discussed above. The very same states turn out to be central in gauging the anomaly free discrete (as well as continuous) symmetries that stabilize proton. In particular, we can gauge these symmetries {\em without} introducing any new fields (compared with the TSSM spectrum given in \cite{zura}) charged under the Standard Model gauge group $SU(3)_c\otimes SU(2)_w\otimes U(1)_Y$. Also, these novel gauge symmetries are similar to but not exactly the same as the ``lepton'' and ``baryon'' numbers. Although we will mainly focus on discrete gauge symmetries, we will also discuss continuous 
gauge symmetries (which must be broken on ``distant'' branes) as they are expected to have interesting  phenomenological implications.

{}Next, let us discuss the neutrino mass problem in the TeV-scale brane world scenario. One of the key observations here is that we no longer have the standard ``see-saw'' mechanism \cite{see-saw}. This is because there is no large (of order $M_{\small{GUT}}$) mass scale in the theory. There is, however, a large {\em distance} scale in the theory, in particular, that associated with the large compact dimensions transverse to the $p$-branes. If we assume that the right-handed neutrinos are ``bulk'' fields propagating in these large dimensions, then small (with the correct order of magnitude) neutrino masses can be generated via the higher dimensional mechanism proposed in \cite{neutrino}. In this mechanism the left-handed neutrinos (which are localized on the $p$-branes) couple to the ``bulk'' right-handed neutrinos, which, in particular, includes couplings to the corresponding tower of Kaluza-Klein states. Then Dirac neutrino masses are generated with $m_\nu\sim M_s^2/M_P$ \cite{neutrino}, where $M_P$ is the four dimensional Planck scale. 

{}The mechanism of \cite{neutrino}, however, crucially depends on the
assumption that the lepton number is conserved. Let us be more precise
here. Note that {\em a priori} we can have a dimension 5 operator
$LLH_+H_+$, where $L$ is the $SU(2)_w$ doublet containing a left-handed
neutrino and the corresponding charged lepton (we are suppressing the
flavor indices), and $H_+$ is the electroweak Higgs doublet with the
hypercharge $+1$. This dimension 5 operator is suppressed by $1/M_s$ only,
and upon the electroweak symmetry breaking would generate Majorana neutrino
masses of order $M_{ew}^2/M_s$ which are obviously unacceptable. (Note that
this is precisely the operator effectively generated by the old ``see-saw''
mechanism except in that case it was suppressed by $1/M_{\small{GUT}}$
giving rise to correct neutrino masses.)  
For the mechanism of \cite{neutrino} as well as for {\em any other} mechanism of neutrino mass generation in TeV-scale brane world scenario to work it is therefore required that this dimension 5 operator be absent (or further suppressed). In \cite{neutrino} this was achieved by assigning the global lepton quantum numbers to the left- and right-handed neutrinos. However, as we already discussed above, we do not expect any global symmetries in the brane world framework. Then we either have to gauge the lepton number, or gauge a discrete subgroup of $U(1)_L$. Both possibilities are non-trivial to realize. Gauging $U(1)_L$ in a straightforward fashion runs into the difficulty of having anomalies as well as into the necessity of eventually breaking $U(1)_L$ on a ``distant'' brane. The latter difficulty is absent if we just gauge a discrete symmetry which can forbid the operator $LLH_+H_+$. It then immediately follows that this discrete symmetry cannot be isomorphic to ${\bf Z}_2$. It is still non-trivial to show that such a discrete symmetry exists, and that it is anomaly free. Here we should point out that if one wishes to generate Majorana neutrino masses, it appears unavoidable to break either the continuous gauge $U(1)_L$ symmetry or its discrete remnant on a ``distant'' brane.

{}In the following we will show that the discrete gauge symmetry which solves the proton decay problem in the TSSM also forbids the dimension 5 operator $LLH_+H_+$ thus killing two birds with one stone. In fact, this is a non-trivial statement. For instance, the ${\bf Z}_3$ ``Generalized Baryon Parity'' symmetry introduced in \cite{IR} alone cannot do the job: the operator $LLH_+H_+$ conserves the corresponding ${\bf Z}_3$ charges. In fact, in \cite{IR} it was shown (in the assumption of the MSSM spectrum plus right-handed neutrinos but no other states charged under $SU(3)_c\otimes SU(2)_w\otimes U(1)_Y$) that this symmetry is the only {\em possible}\footnote{To show that this discrete symmetry is indeed completely anomaly free is, however, a more non-trivial task. In our model this discrete symmetry can indeed be gauged consistently.} anomaly free discrete ${\bf Z}_N$ symmetry which can suppress dimension 5 operators leading to too rapid proton decay and allow for the neutrino masses via the operator $LLH_+H_+$ in the old scenario with $M_s\sim M_{\small{GUT}}$. In the following we will point out that this might allow to distinguish between the high {\em vs.} low $M_s$ scenarios in the collider experiments.

{}Finally, let us mention some of the ``bonuses'' that also arise in the TSSM. Thus, it turns out that the mass hierarchy between up and down quarks ({\em e.g.}, $t$ {\em vs.} $b$) can be explained without appealing to large $\tan\beta$. Also, the $\mu$-term for the electroweak Higgs doublets (as well as for the new states) can be generated.

{}The rest of this paper is organized as follows. In section II we briefly review the TSSM proposed in \cite{zura}. In section III we discuss gauging discrete as well as continuous symmetries in the TSSM, and possible solutions to the anomaly cancellation requirements. In section IV we discuss various phenomenological implications of the TSSM. In particular, we show how the proton decay problem as well as the neutrino mass generation problem can be solved. We also discuss other phenomenological implications of the TSSM including possible signatures for the upcoming collider experiments. In section V we discuss possible brane world (that is, string theory) embeddings of the TSSM. This section contains various string theoretic discussions and can be skipped without breaking the flow of the rest of the paper. In section VI we briefly summarize our results and discuss some open questions. Some details concerning the construction of anomaly free discrete and continuous gauge symmetries are relegated to appendix A. 

\section{The TSSM}

{}In this section we briefly review the TSSM proposed in \cite{zura}. 
The gauge group of this model is the same as in the MSSM, that is, $SU(3)_c\otimes SU(2)_w \otimes U(1)_Y$. The light spectrum\footnote{By the light spectrum we mean the states which are massless before the supersymmetry/electroweak symmetry breaking.} of the model is ${\cal N}=1$ supersymmetric, and along with the vector superfields transforming in the adjoint
of $SU(3)_c\otimes SU(2)_w \otimes U(1)_Y$ we also have the following chiral superfields
(corresponding to the matter and Higgs particles):
\begin{eqnarray}
 && Q_i=3\times ({\bf 3},{\bf 2})(+1/3)~,~~~
 D_i=3\times ({\overline {\bf 3}},{\bf 1})(+2/3)~,~~~
 U_i=3\times ({\overline {\bf 3}},{\bf 1})(-4/3)~,\nonumber\\
 && L_i=3\times ({\bf 1},{\bf 2})(-1)~,~~~E_i=3\times ({\bf 1},{\bf 1})(+2)~,~~~
 N_i=3\times ({\bf 1},{\bf 1})(0)~,\nonumber\\
 && H_+=({\bf 1},{\bf 2})(+1)~,~~~H_-=({\bf 1},{\bf 2})(-1)~,\nonumber\\
 && F_+=({\bf 1},{\bf 1})(+2)~,~~~F_-=({\bf 1},{\bf 1})(-2)~.\nonumber
\end{eqnarray}
Here the $SU(3)_c\otimes SU(2)_w$ quantum numbers are given in bold font, whereas the $U(1)_Y$ hypercharge is given in parentheses. The three generations $(i=1,2,3)$ of quarks 
and leptons are given by $Q_i,D_i,U_i$ respectively $L_i,E_i$ (we have also included the chiral superfields $N_i$ corresponding to the right-handed neutrinos), whereas $H_\pm$ correspond to the electroweak Higgs doublets. Note that the chiral superfields $F_\pm$ are {\em new}: they were not present in the MSSM. 

{}To describe the massive spectrum of the TSSM let us introduce some notations. In the following we will use the collective notation $\Phi\equiv (V,H_+,H_-,F_+,F_-)$, where $V$ stands for the ${\cal N}=1$ vector superfields transforming in the adjoint of $SU(3)_c\otimes SU(2)_w \otimes U(1)_Y$. Now consider a straightforward lifting of the ${\cal N}=1$ superfields $\Phi$ to ${\cal N}=2$ superfields ${\widetilde \Phi}$. In this process the ${\cal N}=1$ vector
multiplets $V$ are promoted to ${\cal N}=2$ vector multiplets ${\widetilde V}=V\oplus\chi$,
where $\chi$ stands for the ${\cal N}=1$ chiral superfields transforming in the adjoint of $SU(3)_c\otimes SU(2)_w \otimes U(1)_Y$. Similarly, the ${\cal N}=1$ chiral supermultiplets 
$H_\pm,F_\pm$ are promoted to the corresponding ${\cal N}=2$ hypermultiplets which differ from the former by the additional ${\cal N}=1$ chiral supermultiplets of opposite chirality but the same gauge quantum numbers. We will refer to these additional ${\cal N}=1$ chiral superfields
as $H^\prime_\pm,F^\prime_\pm$. Then we can write 
the collective notation ${\widetilde \Phi}$ for the ${\cal N}=2$ superfields as
${\widetilde \Phi}=\Phi\oplus\Phi^\prime$, where
$\Phi^\prime\equiv (\chi,H^\prime_+,H^\prime_-,F^\prime_+,F^\prime_-)$.

{}The massive spectrum of the TSSM contains Kaluza-Klein (KK) states. These
states correspond to compact $p-3$ dimensions inside of the D$p$-branes
($p=4$ or 5) on which the gauge fields are localized. The heavy KK levels
are populated by ${\cal N}=2$ supermultiplets with the quantum numbers
given by ${\widetilde \Phi}$. The exact massive KK spectrum is not going to
be important up until section V where we will discuss it in more
detail. Here we would like to point out some of the features of the model
which are going to be relevant for discussions in this section as well as
sections III and IV. Note that the massless superfields $V,H_\pm,F_\pm$
have heavy KK counterparts. We can think about these states together with
the corresponding heavy KK modes as arising upon compactification of a
$p+1$ dimensional theory on a $p-3$ dimensional compact space (whose
precise definition will be given in section V) with the volume
$V_{p-3}$. On the other hand, the massless superfields
$Q_i,D_i,U_i,L_i,E_i,N_i$ do not possess heavy KK counterparts
corresponding to these $p-3$ dimensions\footnote{Actually, here we have a
choice of allowing the chiral superfields $N_i$ (corresponding to
right-handed neutrinos) to have such KK counterparts. This would not affect
the discussion of the gauge coupling unification in \cite{zura} as these
states are neutral under $SU(3)_c\otimes SU(2)_w \otimes U(1)_Y$.}. That
is, the corresponding fields are truly $3+1$ dimensional, and do not
descend from a higher $p+1$ dimensional theory\footnote{{\em A priori} here
we could allow up to two (see \cite{zura} for details) of the generations
to have heavy KK counterparts. This does not spoil the unification
prediction at one loop as each generation of quarks and leptons falls into
complete $SU(5)$ multiplets. However, for the reasons that will become
clear in the following, in this paper we will focus on the case described
above. In particular, this restriction appears to be necessary for our
discussion of discrete gauge symmetries to go through. Also,
in the following we will assume that the $N_i$ superfields have KK
excitations in directions {\em transverse} to the $p$-branes. This,
however,  does not
affect our discussion of gauge coupling unification.}. 
(A concrete mechanism for localizing these fields in this way will be discussed in the brane world context in section V.)        

{}Next, let us briefly discuss the gauge coupling unification in the TSSM. Here we will skip all the details, which can be found in \cite{zura}, and simply state the results. Let $\alpha_a$, $a=1,2,3$, be the $U(1)_Y$, $SU(2)_w$ and $SU(3)_c$
gauge couplings, respectively. (The standard normalization $\alpha_1={5\over 3}\alpha_Y$
is understood.) Let us take the LEP data for the gauge couplings $\alpha_a(\mu)$ at some low energy scale $\mu$ (say, $\mu=M_Z$, where $M_Z$ is the $Z$-boson mass). Then the gauge coupling unification in this model is as precise (at one loop\footnote{Here the following remark is in order. The unified gauge coupling $\alpha$ in the TSSM is small. Thus, for instance, for $M_s\simeq 10~{\mbox{TeV}}$ we have $\alpha\simeq 1/37.5$, so naively it might seem that the perturbation theory is valid. However, as explained in detail in \cite{zura}, the true loop expansion parameter is enhanced by a factor proportional to the number of the heavy KK states which is large. (This enhancement is analogous to that in large $N$ gauge theories \cite{thooft}.) To be specific, let us consider a concrete example where $p=5$ so that the gauge bosons are localized on D5-branes. Let the two dimensions inside of the D5-branes be compactified on the $(S^1\otimes S^1)/{\bf Z}_2$ orbifold, where ${\bf Z}_2$ reflects both coordinates corresponding to the two circles whose radii we will choose identical and equal $R$. Then (in the particular subtraction scheme used in \cite{dien} and also mentioned in \cite{zura}) we have $RM_s\simeq 6.1$. The loop expansion parameter is given by $\lambda_s N$, where $\lambda_s=(\alpha/4\pi)(RM_s)^2\simeq 0.08$, and $N$ is the number of D5-branes. For $N{\ \lower-1.2pt\vbox{\hbox{\rlap{$>$}\lower5pt\vbox{\hbox{$\sim$}}}}\ }10$ the loop expansion parameter is no longer small. For smaller values of $N$, however, the perturbation theory can still be valid. At any rate, as explained in \cite{zura}, the one-loop corrections to the gauge couplings are expected to be dominant even if $\lambda_s N\sim 1$.}) as in the MSSM, and the unification scale $M_s$ can be in the TeV range (provided that the mass scale of the superfields $F_\pm$ is around that of the electroweak Higgs doublets $H_\pm$). (The unification scale depends on the volume $V_{p-3}$ which is assumed to be (relatively) large, that is, $V_{p-3}/(2\pi)^{p-3}\gg 1/M_s^{p-3}$.) The lowering of the unification scale here occurs 
along the lines of \cite{dien} due to the power-like running of the gauge couplings above the KK threshold scale \cite{TV}.

{}Alternatively, we can assume that at some scale $M_s$ (which we will identify with the string scale in the brane world picture) the gauge couplings are the same and equal the unified gauge coupling $\alpha$, and compute the one-loop corrections to the gauge couplings at some low energy scale $\mu\ll1/R<M_s$. Then we can see whether the one-loop corrected gauge couplings at the energy scale $\mu$ agree with the low energy experimental data. Using the results of \cite{TV}, the low energy gauge couplings at one loop are given by\cite{zura}:
\begin{equation}\label{running1}
 \alpha^{-1}_a (\mu) = \alpha^{-1} +
 {{\widehat b}_a\over 2\pi}\log\left({M_s\over \mu}\right)
 +{{\widetilde b}_a\over 2\pi} \eta_p (RM_s)^{p-3} 
 -{{\widetilde b}_a\over 4\pi} \log\left(R M_s\right)~,
\end{equation}
where we are assuming that $(RM_s)^{p-3}\gg 1$. Here the logarithmic contribution is due to the light (massless) modes, whereas the $\mu$-independent terms are due to the heavy KK {\em thresholds}. The $\beta$-function coefficients ${\widehat b}_a$ correspond to the ${\cal N}=1$
supersymmetric light modes (${\widehat b}_1={39\over 5}$, ${\widehat b}_2=1$, ${\widehat b}_3=-3$). The heavy KK thresholds are proportional to the corresponding ${\cal N}=2$
$\beta$-function coefficients ${\widetilde b}_a$ computed for the superfields ${\widetilde \Phi}$ (recall that the heavy KK levels are populated by ${\cal N}=2$ supermultiplets with the ${\widetilde \Phi}$ quantum numbers):
\begin{equation}\label{N=2}
 {\widetilde b}_1={18\over 5}~,~~~{\widetilde b}_2=-2~,~~~{\widetilde b}_3=-6~.
\end{equation}
Finally, the coefficient $\eta_p$ is a (strictly speaking subtraction scheme dependent) numerical factor of order one whose precise value is not going to be important here (see \cite{zura}
for details).

{}Note that the dominant contributions to the differences between the gauge couplings $\alpha_a(\mu)$ come from the heavy KK thresholds provided that $(RM_s)^{p-3}\gg 1$. In particular, in the leading approximation we can neglect the logarithmic terms. The TSSM predictions for the low energy gauge couplings $\alpha_a(\mu)$ then agree with the experimental data just as precisely (at one loop) as in the MSSM provided that the unification scale $M_s$ and the unified gauge coupling $\alpha$ are given by
\begin{eqnarray}\label{uniA}
 &&\eta_p (RM_s)^{p-3} 
 \approx  \log \left({M_{\small{GUT}}\over M_s}\right)~,\\
 \label{uniA1}
 &&\alpha^{-1}\approx\alpha^{-1}_{\small{GUT}}+{3\over 2\pi} \log\left( {M_{\small{GUT}}
 \over M_s}\right)~,
\end{eqnarray}
where $\alpha_{\small{GUT}}\approx1/24$ and $M_{\small{GUT}}\approx 2\times 10^{16}~{\mbox{GeV}}$ are the unified gauge coupling respectively the unification scale in the context of the MSSM \cite{gut}. The reason why the unification of gauge couplings in the TSSM is as precise (at one loop) as in the MSSM is that  
\begin{equation}\label{nu}
  {{\widetilde b}_a-{\widetilde b}_b
 \over b_a-b_b}=1~~~\forall~a\not=b~.
\end{equation}  
Here $b_a$ are the corresponding $\beta$-function coefficients in the MSSM:
\begin{equation}
 b_1={33\over 5}~,~~~b_2=1~,~~~b_3=-3~.
\end{equation}
Here we would like to stress that the new states $F_\pm$ that we have introduced in the TSSM are crucial for the unification prediction: the condition (\ref{nu}) would {\em not} be satisfied (which would ruin the unification prediction) if we did not include these states accompanied by their heavy KK counterparts.

\section{Discrete and Continuous Gauge Symmetries}

{}In this section we will discuss discrete and continuous symmetries which can be consistently gauged in the TSSM reviewed in the previous section. The consistency conditions for gauging such symmetries are dictated by the requirement that they be anomaly free. 
In the next section we will show that some of these gauge symmetries stabilize proton in the TSSM, and are also required for successful generation of small neutrino masses. 

{}Let us start with the discrete gauge symmetries. Here we will confine our attention to Abelian finite discrete groups, and for simplicity our discussion will be in the context of ${\cal N}=1$ supersymmetric theories. Thus, consider a theory with some assignment of ${\bf Z}_N$ discrete charges. The anomaly cancellation conditions for such symmetries were discussed in \cite{IR1}
(also see, {\em e.g.}, \cite{inst}). These conditions are analogous to those for a $U(1)$ gauge symmetry. In fact, a ${\bf Z}_N$ discrete gauge symmetry can be thought of as follows. Consider a theory with some matter charged under an anomaly free $U(1)$ gauge symmetry. Let all the $U(1)$ charge assignments be integer. Consider now adding a pair of chiral superfields which are neutral under all the other gauge subgroups but carry $+N$ and $-N$ charges under the above $U(1)$. Suppose there is a flat direction along which these chiral superfields acquire non-zero VEVs. Then the $U(1)$ gauge symmetry is broken down to its ${\bf Z}_N$ subgroup. This ${\bf Z}_N$ is then an anomaly free discrete gauge symmetry. In the above approach the anomalies for the ${\bf Z}_N$ gauge symmetry mimic those for the original $U(1)$ gauge symmetry (except that the former anomalies are only defined ``modulo $N$''). Thus, we have the ${\mbox{Tr}}({\bf Z}_N^3)$ anomaly as well as the mixed ${\mbox{Tr}}({\bf Z}_N)$ gravitational anomaly. If there are {\em non-Abelian} gauge subgroups $\bigotimes_i G_i$ in the theory, one also needs to consider the mixed ${\mbox{Tr}}(G_i^2 {\bf Z}_N)$ non-Abelian gauge anomalies. Finally, if there are additional {\em Abelian} subgroups $\bigotimes_a U(1)_a$, then we must also consider the mixed ${\mbox{Tr}}(U(1)_a U(1)_b {\bf Z}_N)$ and ${\mbox{Tr}}(U(1)_a {\bf Z}^2_N)$ gauge anomalies. The last two anomalies are somewhat tricky compared with the rest of the anomalies. The reason is that to compute them one is required to know the massive spectrum of the theory. More concretely, these anomalies depend on details of the parent $U(1)$ breaking \cite{IR1}. Since we are going to attempt to gauge discrete symmetries in the TSSM which contains an Abelian gauge subgroup (namely, $U(1)_Y$), we cannot ignore these anomalies. There is however, a way out of this difficulty. We will explicitly gauge {\em continuous} (that is, $U(1)$) symmetries and make sure that all the anomalies cancel before we break them to the corresponding discrete subgroups. This way we are guaranteed to have an anomaly free discrete gauge symmetry at the end of the day.

{}In order to simplify our task of finding anomaly free gauge symmetries in our model, we can first use some phenomenological constraints to reduce the number of viable possibilities. Thus, in appendix A we briefly review some of the discussions of \cite{IR}\footnote{For earlier works, see, {\em e.g.}, \cite{hall}.} which lead to the conclusion that there are only two distinct ${\bf Z}_N$ symmetries we can consider. Let us be more precise here. First, the ${\bf Z}_N$ symmetries we are going to use in this paper will be ``generation blind'', that is, they have identical actions on all three generations. Second, we must impose the following requirements: ({\em i}) the chiral fermions should have the usual Yukawa couplings with the electroweak Higgs doublets $H_\pm$;  ({\em ii}) the $\mu$-term $\mu H_+ H_-$ should be allowed. These constraints are satisfied by only two inequivalent discrete symmetries (see appendix A for details). We will refer to the generators of these two discrete ${\bf Z}_N$ symmetries as ${\cal L}$ and ${\cal R}$. In fact, we will use the same notation for the corresponding parent $U(1)$'s as well as the $U(1)$ charges. As we have already mentioned above, we will explicitly construct anomaly free parent $U(1)$'s to make sure that their discrete subgroups are also anomaly free. Let us, therefore, discuss the $U(1)_{\cal L}$ and $U(1)_{\cal R}$ charge assignments for the fields in the TSSM. 

\begin{center}
{\em The $U(1)_{\cal L}$ Symmetry}
\end{center}     

{}The $U(1)_{\cal L}$ charge assignments are given by (since $U(1)_{\cal L}$ (as well as $U(1)_{\cal R}$) symmetry is ``generation blind'', in the following we will omit the flavor indices):
\begin{eqnarray}
 &&{\cal L}_Q =0~,~~~{\cal L}_D =0~,~~~{\cal L}_U =0~,\nonumber\\
 &&{\cal L}_L =+1~,~~~{\cal L}_E =-1~,~~~{\cal L}_N =-1~,\nonumber\\
 &&{\cal L}_{H_+} =0~,~~~{\cal L}_{H_-} =-3~,\nonumber\\
 &&{\cal L}_{F_+} =+3~,~~~{\cal L}_{F_-} =0~,\nonumber\\
 &&{\cal L}_S = +3~.\nonumber
\end{eqnarray} 
Note that the charge assignments for the fields $Q,D,U,L,E,N$ follow\footnote{Note that including the right-handed neutrinos is required by the anomaly cancellation conditions in this case.} from the considerations of appendix A. The charge assignments for the fields $H_\pm,F_\pm,S$ are then uniquely\footnote{\label{foot}Here we should point out that there is one other solution to the anomaly cancellation conditions where all the fields except for $H_\pm,F_\pm$ have the same $U(1)_{\cal L}$ charge assignments
as above, and  ${\cal L}_{H_+} =-3$, ${\cal L}_{H_-}=0$,
${\cal L}_{F_+} =0$, ${\cal L}_{F_-} =+3$. {\em A priori} both assignments are perfectly consistent, and might even look equivalent as far as phenomenology is concerned. 
However, as we will point out in the next section, the first assignment gives rise to the desired hierarchy between the up and down quark masses, whereas the hierarchy is reversed for the second assignment. We will therefore focus on the first assignment throughout this paper, albeit it is completely straightforward to repeat the subsequent discussions for the second possibility.} 
fixed by the requirement of anomaly cancellation. In particular, the ${\mbox{Tr}}(U(1)_{\cal L})$, ${\mbox{Tr}}(U(1)^3_{\cal L})$, ${\mbox{Tr}}(SU(2)_w^2 U(1)_{\cal L})$,
${\mbox{Tr}}(SU(3)_c^2 U(1)_{\cal L})$, as well as ${\mbox{Tr}}(U(1)_Y^2 U(1)_{\cal L})$ and ${\mbox{Tr}}(U(1)_Y U(1)_{\cal L}^2)$ anomalies cancel in this model. The anomaly cancellation requirement is the reason for introducing the additional field $S$. This field, however, is {\em neutral} under the Standard Model gauge group. Note that the ${\cal L}$ charges for the fields 
$Q,D,U,L,E,N$ are the same as the familiar lepton number. However, ${\cal L}$ is {\em not} exactly the gauged version of the lepton number because of the charge assignments for the fields $H_\pm,F_\pm,S$.

\begin{center}
{\em The $U(1)_{\cal R}$ Symmetry}
\end{center}     

{}The $U(1)_{\cal R}$ charge assignments are given by:
\begin{eqnarray}
 &&{\cal R}_Q =0~,~~~{\cal R}_D =+1~,~~~{\cal R}_U =-1~,\nonumber\\
 &&{\cal R}_L =0~,~~~{\cal R}_E =+1~,~~~{\cal R}_N =-1~,\nonumber\\
 &&{\cal R}_{H_+} =+1~,~~~{\cal R}_{H_-} =-1~,\nonumber\\
 &&{\cal R}_{F_+} =+2~,~~~{\cal R}_{F_-} =-2~,\nonumber\\
 &&{\cal R}_S = 0~.\nonumber
\end{eqnarray} 
Note that the charge assignments for the fields $Q,D,U,L,E,N$ follow from the considerations of appendix A. The charge assignments for the fields $H_\pm,F_\pm,S$ are then uniquely 
fixed by the requirement of anomaly cancellation (see below) - all 
anomalies cancel in this model.
Note that the field $S$ is neutral under $U(1)_{\cal R}$ so it need not be introduced when discussing $U(1)_{\cal R}$ without $U(1)_{\cal L}$. However, we have kept this field for the later convenience. Note that the ${\cal R}$ charges are the same (up to a sign convention) as (twice) the third component of the right-handed weak isospin\footnote{Thus, consider breaking of the Pati-Salam gauge group $SU(4)_c\otimes SU(2)_w\otimes SU(2)_R$ to $SU(3)_c\otimes SU(2)_w \otimes U(1)_Y$. The fields $Q$ and $L$ are $SU(2)_R$ singlets. The fields $(D,U)$, $(E,N)$ and $(H_+,H_-)$ form $SU(2)_R$ doublets. Finally, the fields $F_\pm$ (together with a singlet, which, for instance, can be identified with $S$) come from a triplet (adjoint) representation of $SU(2)_R$.}.

{}Note that the fields $H_\pm,F_\pm,S$ do not actually contribute to the anomalies for $U(1)_{\cal R}$. The charge assignments for $H_\pm$ are still fixed by the considerations of appendix A, however, the $F_\pm$ charge assignments are not. The above charge assignments are uniquely fixed if we consider gauging {\em both} $U(1)_{\cal L}$ and $U(1)_{\cal R}$ at the same time. (This will be important in the next section as discrete subgroups of $U(1)_{\cal L}$ or $U(1)_{\cal R}$ by themselves do not stabilize proton.) In fact, it is not difficult to check that with the above $U(1)_{\cal L}$ and $U(1)_{\cal R}$ assignments all anomalies cancel if we gauge $U(1)_{\cal L}\otimes U(1)_{\cal R}$. In particular, the mixed ${\mbox{Tr}}(U(1)_{\cal L} U(1)_{\cal R}^2)$, ${\mbox{Tr}}(U(1)^2_{\cal L} U(1)_{\cal R})$ and ${\mbox{Tr}}(U(1)_Y U(1)_{\cal L} U(1)_{\cal R})$ anomalies all cancel.

{}From the above discussion it should be clear that we can actually gauge infinitely many anomaly free $U(1)$'s in the TSSM. Thus, consider a theory with the gauge group $SU(3)_c\otimes SU(2)_w \otimes U(1)_Y\otimes \bigotimes_a U(1)_{{\cal F}_a}$, where
\begin{equation}
 {\cal F}_a =\alpha_a {\cal L}+\beta_a {\cal R} +\gamma_a Y~,
\end{equation}  
and the $U(1)_{{\cal F}_a}$ charges are determined by the exact same linear combination
via the ${\cal L}, {\cal R}, Y$ charges. Here $\alpha_a,\beta_a,\gamma_a$ are arbitrary coefficients. It is clear that all anomalies cancel in such a theory. Note, however, that not all of these additional $U(1)$'s are helpful in forbidding a given process. More precisely, together with $U(1)_Y$ only two additional (linearly independent) symmetries are useful from this viewpoint as the conservation laws due to all the other $U(1)$'s are subsumed in those due to the first three $U(1)$'s.

{}To illustrate the above discussion, let us consider the linear combination ${\cal Z}\equiv Y-{\cal R}$. It is not difficult to see that ${\cal Z}$ is nothing but what is commonly referred to as the $B-L$ charge (baryon number minus lepton number). It is anomaly free and can be gauged together with $U(1)_{\cal L},U(1)_{\cal R}$. However, it will not impose any additional constraints on allowed couplings on top of those already implied by the $U(1)_{\cal L}$ plus $U(1)_{\cal R}$ conservation (here we are taking into account that $U(1)_Y$ must be conserved). For our purposes, therefore, it will suffice to gauge at most $U(1)_{\cal L}$ and $U(1)_{\cal R}$ together. As we will see in the next section, however, gauging just some of their linear combinations may also suffice for our purposes. Here we would like to discuss two of these combinations in more detail for the later convenience. These combinations are given by ${\cal X} \equiv {\cal L}+ {\cal R}$ and ${\cal Y} \equiv {\cal L}- {\cal R}$.

\begin{center}
{\em The $U(1)_{\cal X}$ Symmetry}
\end{center}     

{}The $U(1)_{\cal X}$ charge assignments are given by:
\begin{eqnarray}
 &&{\cal X}_Q =0~,~~~{\cal X}_D =+1~,~~~{\cal X}_U =-1~,\nonumber\\
 &&{\cal X}_L =+1~,~~~{\cal X}_E =0~,~~~{\cal X}_N =-2~,\nonumber\\
 &&{\cal X}_{H_+} =+1~,~~~{\cal X}_{H_-} =-4~,\nonumber\\
 &&{\cal X}_{F_+} =+5~,~~~{\cal X}_{F_-} =-2~,\nonumber\\
 &&{\cal X}_S = +3~.\nonumber
\end{eqnarray} 

\begin{center}
{\em The $U(1)_{\cal Y}$ Symmetry}
\end{center}     

{}The $U(1)_{\cal Y}$ charge assignments are given by:
\begin{eqnarray}
 &&{\cal Y}_Q =0~,~~~{\cal Y}_D =-1~,~~~{\cal Y}_U =+1~,\nonumber\\
 &&{\cal Y}_L =+1~,~~~{\cal Y}_E =-2~,~~~{\cal Y}_N =0~,\nonumber\\
 &&{\cal Y}_{H_+} =-1~,~~~{\cal Y}_{H_-} =-2~,\nonumber\\
 &&{\cal Y}_{F_+} =+1~,~~~{\cal Y}_{F_-} =+2~,\nonumber\\
 &&{\cal Y}_S = +3~.\nonumber
\end{eqnarray} 

{}It is now straightforward to obtain anomaly free ${\bf Z}_N$ discrete gauge symmetries. Thus, let us take a $U(1)_{{\cal F}_a}$ gauge symmetry constructed as discussed above. Introduce two chiral superfields $S^\prime_a$ and $S^{\prime\prime}_a$ which are neutral under all the gauge symmetries except for $U(1)_{{\cal F}_a}$ itself, and carry $+N$ respectively $-N$ $U(1)_{{\cal F}_a}$ charges. (Note that this does not change any of the anomalies.) Suppose that there is a flat direction along which $S^\prime_a,S^{\prime\prime}_a$ acquire non-zero VEVs. Then the $U(1)_{{\cal F}_a}$ symmetry is broken down to its ${\bf Z}_N$ subgroup, and we will refer to the corresponding discrete gauge symmetry as ${\widetilde {\cal F}}_{a,N}$. In the next section we will make use of ${\bf Z}_3$ discrete gauge symmetries ${\widetilde {\cal L}}_3,{\widetilde {\cal R}}_3,{\widetilde {\cal X}}_3,{\widetilde {\cal Y}}_3$. Here we give the corresponding ${\bf Z}_3$ charges for the fields
$Q,D,U,L,E,N,H_\pm,F_\pm,S$ for the later convenience (the $({\widetilde {\cal L}}_3,{\widetilde {\cal R}}_3,{\widetilde {\cal X}}_3,{\widetilde {\cal Y}}_3)$ charges are given in parentheses):   
\begin{eqnarray}
 &&Q:~(0,0,0,0)~,~~~D:~(0,+1,+1,-1)~,~~~U:~(0,-1,-1,+1)~,\nonumber\\
 &&L:~(+1,0,+1,+1)~,~~~E:~(-1,+1,0,+1)~,~~~N:~(-1,-1,+1,0)~,\nonumber\\
 &&H_+:~(0,+1,+1,-1)~,~~~H_-:~(0,-1,-1,+1)~,\nonumber\\
 &&F_+:~(0,-1,-1,+1)~,~~~F_-:~(0,+1,+1,-1)~,\nonumber\\
 &&S_1,S_2:~(0,0,0,0)~.\nonumber
\end{eqnarray} 
Here we have chosen to normalize these discrete charges to integers so that they have to be conserved modulo 3. The singlets $S_1,S_2$ are linear combinations of the original chiral superfields $S,S^\prime,S^{\prime\prime}$. (One linear combination of the latter has been eaten in the super-Higgs mechanism\footnote{Note that in the ${\widetilde {\cal R}}$ case we can have only one singlet which is a linear combination of $S^\prime,S^{\prime\prime}$.}.)

{}Before we end this section, a few remarks are in order. The ${\bf Z}_N$ discrete gauge symmetry ${\widetilde {\cal R}}_N$ was considered in \cite{IR} in the context of the MSSM with the right-handed neutrinos (that is, with chiral superfields $N_i$). There the chiral superfields $F_\pm$ are not present. Nonetheless, the ${\widetilde {\cal R}}_N$ symmetry is anomaly free 
since the $F_\pm$ states do not contribute into the $U(1)_{\cal R}$ anomaly. In \cite{IR} the
${\widetilde {\cal L}}_3,{\widetilde {\cal X}}_3,{\widetilde {\cal Y}}_3$ symmetries (referred to as ${\bf Z}_3$ generalized lepton, matter and baryon parities, respectively) were also discussed but without the $F_\pm$ states. However, the latter are crucial for the anomaly cancellation. Thus, consider the ${\widetilde {\cal L}}_3$ symmetry. The corresponding $U(1)_{\cal L}$ has mixed ${\mbox{Tr}}(U(1)_{\cal L} U(1)_Y^2)$ and ${\mbox{Tr}}(U(1)^2_{\cal L} U(1)_Y)$ anomalies if we drop the $F_\pm$ states. In fact, to gauge $U(1)_{\cal L}$ the states (which upon $U(1)_{\cal L}$ breaking become heavy) analogous to $F_\pm$ were introduced in \cite{IR}. Similar remarks apply to the ${\widetilde {\cal X}}_3,{\widetilde {\cal Y}}_3$ symmetries\footnote{To gauge $U(1)_{\cal Y}$ certain states other than $F_\pm$ were introduced in \cite{IR}, which, in particular, included states with fractional electric charges. Here we see that the $U(1)_{\cal Y}$ as well as any other combination of ${\cal L}$ and ${\cal R}$ can be easily gauged if we introduce the $F_\pm$ states.}. Another important point is that in \cite{IR} the $H_\pm$ states were neutral under ${\cal L}$. This then ultimately requires introduction of additional electroweak doublets (with non-zero ${\cal L}$ charges) to be able to gauge $U(1)_{\cal L}$. If one tries to naively gauge the ${\widetilde {\cal L}}_N$ discrete symmetry (without introducing additional doublets) using the discrete anomaly cancellation conditions of \cite{IR1}, the solution then exists only for $N=3$. However, as we have already pointed out, this solution is not guaranteed to be anomaly free as the mixed ${\mbox{Tr}}({\bf Z}_3 U(1)_Y^2)$ and ${\mbox{Tr}}({\bf Z}_3^2 U(1)_Y)$ anomalies cannot be checked this way. Similar remarks apply to the ${\widetilde {\cal X}}_3,{\widetilde {\cal Y}}_3$ symmetries as well. The construction of this section guarantees that the corresponding discrete gauge symmetries are anomaly free. Moreover, it is possible to obtain ${\widetilde {\cal L}}_N,{\widetilde {\cal R}}_N,{\widetilde {\cal X}}_N,{\widetilde {\cal Y}}_N$ (as well as all the other linear combinations of ${\cal L}$ and ${\cal R}$) for any $N$ (and not just for $N=3$)\footnote{However, the required Yukawa couplings of quarks and leptons to the electroweak Higgs doublets exist only for ${\widetilde {\cal L}}_3,{\widetilde {\cal X}}_3,{\widetilde {\cal Y}}_3$ as well as ${\widetilde {\cal R}}_N$ with arbitrary $N$ - see appendix A and subsection D of section IV.}. This is due to the proper $U(1)_{\cal L}$ charge assignments for the fields $H_\pm$ as well as $F_\pm$.     

\section{Proton Stability, Neutrino Masses, and All That}

{}In this section our main focus will be on proton stability and neutrino masses.
However, we will also  discuss some other phenomenological implications of the TSSM 
such as possible signatures for the upcoming collider experiments.

\subsection{Proton Stability}

{}In this subsection we will address the issue whether the discrete gauge symmetries discussed in the previous section can suppress dangerous higher dimensional operators potentially leading to too rapid proton decay. Let us, however, first briefly review the essence of the proton stabilization problem\footnote{Everything we will say about proton can be straightforwardly generalized to neutron as well.}.

{}Proton is a fermion with the baryon number $+1$, and kinematically it can only decay into an odd number of leptons\footnote{Here we also mean antileptons. Note that the $\tau$-lepton is excluded from the final decay products as its too heavy.} plus, possibly, some light mesons.  
Therefore, any channel through which proton can possibly decay should violate the baryon number by one unit, and the lepton number by an odd integer. This limits possible {\em effective} operators which can be responsible for proton decay. These effective operators have at least dimension 6, but they are not necessarily suppressed by $1/M_s^{d-4}$, where $d$ is the dimension of a given operator. This is because such effective operators can, among other ways, arise via exchange of scalar superpartners (sparticles) coupled to the usual Standard Model states (quarks and leptons). An effective operator arising this way is then suppressed by some powers of $1/M_{\small{SUSY}}$ (and, possibly, $1/M_s$), where $M_{\small{SUSY}}$ is the typical mass scale of the sparticle states. 

{}Let us write down all {\em a priori} possible dimension 3 and 4 operators in the TSSM which are allowed by the $SU(3)_c\otimes SU(2)_w \otimes U(1)_Y$ gauge symmetry\footnote{For the remainder of this section we will focus on the light spectrum as the heavy KK modes mentioned in section II are not going to be relevant for the subsequent discussions.}. The dimension 3 operators are given by
\begin{eqnarray}\label{dim3}
 \left[\mu H_+H_- +\mu^\prime F_+ F_- + m_i L_i H_+ + m^\prime_i E_i F_- 
 + m_{ij} N_i N_j\right]_F~,
\end{eqnarray}  
whereas the possible dimension 4 operators read
\begin{eqnarray}\label{dim4}
 && \left[d_i N_i H_+ H_- + d^\prime_i N_i F_+F_- + h_{ij} L_i E_j H_- +h^\prime_{ij} L_i N_j  
 H_+~,\right.\nonumber\\
 && f_{ij} Q_i D_j H_- +f^\prime_{ij} Q_i U_j H_+  +\lambda_{ij} L_i L_j F_+ 
 +\lambda^\prime_{ij} E_i N_j F_-~,\nonumber\\
 &&\left. 
       \lambda_{ijk} L_iL_j E_k + \lambda^\prime_{ijk} L_iQ_j D_k +
       \lambda^{\prime\prime}_{ijk} U_iD_j D_k\right]_F~.
\end{eqnarray}
Note that the third and the fourth terms in (\ref{dim3}) give rise to mixing between $L$ and $H_-$ respectively $E$ and $F_+$. They can be eliminated by mass diagonalization as a result of which the first and the second terms in the last line of (\ref{dim4}) are generated. The three 
operators in the last line of (\ref{dim4}) are dangerous (via a graph containing a sparticle exchange) for proton stability if present simultaneously \cite{wein}. Let us focus on these three operators for the moment. Since {\em both} baryon and lepton violating vertices are required for proton decay, proton is stable to this order if at least one of the last two of these three operators is absent. (The first two vertices violate lepton number, whereas the third vertex violates the baryon number.)  Note that both ${\widetilde{\cal L}}_N$ as well as 
${\widetilde{\cal R}}_N$ discrete gauge symmetries do the job of forbidding at least one of these two operators. For instance, ${\widetilde{\cal L}}_3$ forbids the lepton number violating operators in both (\ref{dim3}) (these are $LH_+$, $EF_-$ and $NN$ operators) and (\ref{dim4})
(these are $NH_+H_-$, $NF_+F_-$, $LLF_+$, $ENF_-$, $LLE$ and $LQD$ operators). It does allow the baryon number violating vertex $UDD$, however. On the other hand, 
${\widetilde{\cal R}}_3$ forbids all the lepton as well as baryon number violating operators in (\ref{dim3}) and (\ref{dim4}), and so does ${\widetilde{\cal X}}_3$. Finally, ${\widetilde{\cal Y}}_3$ allows {\em all} lepton number violating terms in (\ref{dim3}) and (\ref{dim4}), but forbids the baryon number violating term $UDD$ in (\ref{dim4}) protecting proton from decaying to this order.

{}Suppressing dimension 4 operators, however, is not sufficient to guarantee proton stability. Thus, consider the simplest dimension 5 operator  
\begin{eqnarray}
 \left[QQQL\right]_F~.
\end{eqnarray}
This operator is suppressed by $M_s$ and leads to proton decay with the rate $\Gamma_p/m_p\sim m^4_p/M^2_s M_{\small{SUSY}}^2$ via a one-loop graph involving a chargino exchange. (Here $m_p$ is the proton mass.) For $M_s$ in the TeV range this operator is disastrous for proton stability if present. In fact, this is true even if $M_s$ is as high as $M_{\small{GUT}}$.
Note that ${\widetilde{\cal R}}_3$ does {\em not} forbid this dimension 5 operator. This is actually the case for all the ${\widetilde{\cal R}}_N$ symmetries. Thus, these symmetries by themselves cannot guarantee proton stability. On the other hand, the ${\widetilde{\cal L}}_3,{\widetilde{\cal X}}_3,{\widetilde{\cal Y}}_3$ symmetries do forbid this dimension 5 operator. This leads us to the conclusion that if the solution to the proton stability problem can be found by considering discrete gauge symmetries, the corresponding symmetry should be 
either ${\widetilde{\cal L}}_N$, or ${\widetilde{\cal L}}_N\otimes {\widetilde{\cal R}}_M$, or a subgroup of the latter. In subsection C we will identify a discrete symmetry that will do the job.
Before we do this, however, we would like to discuss the problem of generating small neutrino masses in the TeV-scale brane world scenario. This will provide additional constraints on the discrete gauge symmetries we are looking for. 

\subsection{Neutrino Masses}

{}There is a key difference between how small neutrino masses are supposed to be generated in scenarios with high $M_s$ (of order $M_{\small{GUT}}$) and low $M_s$ (in the TeV range). In the former case we can have some variant of the ``see-saw'' mechanism \cite{see-saw} via which an effective dimension 5 operator 
\begin{equation}
 \left[LLH_+H_+\right]_F
\end{equation}
is generated. This operator is suppressed by $1/M_s$, so upon the electroweak symmetry breaking small neutrino masses $m_\nu\sim M_{ew}^2/M_s\sim M_{ew}^2/M_{\small{GUT}}$ are generated, and these neutrino masses are in the correct bulk range. Note that this dimension 5 operator is lepton number violating. 

{}On the other hand, the presence of the dimension 5 operator $LLH_+H_+$ in the TeV-scale brane world scenario would be disastrous: it would immediately generate large Majorana masses of order $M_{ew}^2/M_s$ in gross contradiction with the experimental data. Thus, it is mandatory that we forbid this dimension 5 operator in any TeV-scale brane world scenario. It is therefore reasonable to ask whether any of the discrete symmetries in the TSSM we have considered so far can do this job. For concreteness, let us focus on the ${\widetilde{\cal L}}_3,{\widetilde{\cal R}}_3,{\widetilde{\cal X}}_3,{\widetilde{\cal Y}}_3$ symmetries.

{}Thus, the symmetries ${\widetilde{\cal L}}_3,{\widetilde{\cal R}}_3,{\widetilde{\cal X}}_3$ do forbid the above dimension 5 operator. However, the ${\widetilde{\cal Y}}_3$ symmetry allows this operator. This implies that ${\widetilde{\cal Y}}_3$ alone cannot guarantee that large Majorana neutrino masses are not generated in a particular TeV brane world scenario. This is essentially due to the fact that, as we pointed out in the previous subsection, the ${\widetilde{\cal Y}}_3$ symmetry allows the lepton number violation already at dimension 3 and dimension 4 level.

{}We can therefore conclude that we need to find a discrete gauge symmetry that conserves the lepton number at least up to some higher dimensional operator. This implies that to generate small Majorana neutrino masses we must break this symmetry without generating an unacceptably large contribution from the $LLH_+H_+$ operator\footnote{Note that in \cite{neutrino1} a higher dimensional mechanism for generating small Majorana neutrino masses was proposed. For this mechanism to work it is required that the lepton number non-conservation is maximal to begin with. It is then unclear what suppresses the $LLH_+H_+$ operator.}. This can presumably be done by breaking the ``lepton number'' symmetry (or, more precisely, its discrete version) on some ``distant'' brane. At present the precise mechanism of such breaking is not completely clear in the brane world context, so we will leave this question for the future investigations. Note, however, that small Dirac neutrino masses can be generated via the higher dimensional mechanism proposed in \cite{neutrino}. We will discuss possible brane world embeddings of this mechanism 
in section V.      

\subsection{A Conclusive Solution}

{}We are now ready to write down a discrete gauge symmetry that stabilizes proton and at the same time solves the problem of dimension 5 lepton number violating operator discussed in the previous subsection. The symmetry we are proposing here is the ${\widetilde{\cal L}}_3\otimes {\widetilde{\cal R}}_3$ symmetry (which is a ${\bf Z}_3\otimes {\bf Z}_3$ symmetry). Thus, consider the TSSM augmented with this discrete symmetry. For the reader's convenience, here we give the light spectrum of this model with all the quantum numbers under  
$SU(3)_c\otimes SU(2)_w\otimes U(1)_Y\otimes {\widetilde{\cal L}}_3\otimes {\widetilde{\cal R}}_3$ (we omit the singlets which are neutral under all of these symmetries):
\begin{eqnarray}
 && Q_i=3\times ({\bf 3},{\bf 2})(+1/3)[0,0]~,\nonumber\\
 &&D_i=3\times ({\overline {\bf 3}},{\bf 1})(+2/3)[0,+1]~,~~~
 U_i=3\times ({\overline {\bf 3}},{\bf 1})(-4/3)[0,-1]~,\nonumber\\
 && L_i=3\times ({\bf 1},{\bf 2})(-1)[+1,0]~,\nonumber\\
 && E_i=3\times ({\bf 1},{\bf 1})(+2)[-1,+1]~,~~~
 N_i=3\times ({\bf 1},{\bf 1})(0)[-1,-1]~,\nonumber\\
 && H_+=({\bf 1},{\bf 2})(+1)[0,+1]~,~~~H_-=({\bf 1},{\bf 2})(-1)[0,-1]~,\nonumber\\
 && F_+=({\bf 1},{\bf 1})(+2)[0,-1]~,~~~F_-=({\bf 1},{\bf 1})(-2)[0,+1]~.\nonumber
\end{eqnarray}
The $U(1)_Y$ charges are given in parentheses, whereas the ${\widetilde{\cal L}}_3\otimes {\widetilde{\cal R}}_3$ discrete charges (which are conserved modulo 3) are given in square brackets. As we pointed out in section III, all of the above symmetries are anomaly free in this model.

{}The allowed dimension 3 and 4 operators in this model are given by
\begin{eqnarray}\label{dim34}
 && \left[\mu H_+H_- +\mu^\prime F_+ F_- + ~,\right.\nonumber\\
 && \left. 
 h_{ij} L_i E_j H_- +h^\prime_{ij} L_i N_j  
 H_+  +f_{ij} Q_i D_j H_- +f^\prime_{ij} Q_i U_j H_+\right]_F ~,
\end{eqnarray}
which are the same as in the MSSM except for the $\mu^\prime$-term for the new $F_\pm$ states, and the usual Yukawa coupling $LN H_+$ for the right-handed neutrinos.  
As to the dimension 5 operators, we will show in a moment that all dangerous higher dimensional operators (including dimension 5) possibly leading to proton decay are absent in this model, and, as we have already mentioned, the dangerous lepton number violating dimension 5 operator $LLH_+H_+$ is also absent (it is not difficult to see that all the other operators of this type such as $LLH_+H_+H_+H_-$ are also absent). Before we turn to the baryon and lepton number violating higher dimensional operators, however, we would like to discuss dimension 5 operators involving the new $F_\pm$ states. It is straightforward to check that the following dimension 5 operators are allowed in this model:
\begin{eqnarray}\label{dim5}
 && \left[g_{ij} Q_i D_j H_+ F_- + g^\prime_{ij} Q_i U_j H_- F_+ \right]_F ~.
\end{eqnarray} 
upon the electroweak symmetry breaking these operators give rise to
effective dimension 4 operators involving the $F_\pm$ state with the Yukawa
couplings suppressed by a factor of $M_{ew}/M_s$. This implies that the new
states $F_\pm$ are relatively long lived (compared with the Higgsinos as
well as other sparticle states)\footnote{The fact that the $F_\pm$ states
are unstable is a desirable feature from the viewpoint of various
cosmological as well as other constraints.}, and can be expected to be
relatively easily detectable in collider experiments. Note that the
``competing'' processes involving the MSSM states are actually suppressed:
for instance, the (lepton number violating) dimension 5 operator $QUH_-E$
is forbidden by the discrete symmetries of the model. We can therefore
expect distinct signatures of the $F_\pm$ states at the upcoming collider experiments provided that their mass scale is around that of the electroweak Higgs doublets.

{}Finally, let us show that {\em all} higher dimensional operators possibly leading to proton decay in this model are absent. It is not difficult to see (see below) that to show this it suffices to consider effective operators of the following type\footnote{It should be clear that the singlet states neutral under the $SU(3)_c\otimes SU(2)_w\otimes U(1)_Y\otimes {\widetilde{\cal L}}_3\otimes {\widetilde{\cal R}}_3$ symmetry are irrelevant for the following discussion. Also, we do not include the $F_\pm$ states in the possibly dangerous operators as they are too heavy. However, we must consider the $H_\pm$ insertions as the fields $H_\pm$ acquire VEVs resulting in electroweak symmetry breaking.}: 
\begin{equation}\label{higher}
QQQL^{k+1} H_+^k N^m,~~~QQQL^{k+1} H_+^k N^{*m}~. 
\end{equation}
These operators are {\em a priori} allowed for any $k,m$ by the $SU(3)_c\otimes SU(2)_w\otimes U(1)_Y$ symmetry. Note that technically speaking we can even restrict $k+m$ to be an even integer. This is because in the final state we must have an odd number of leptons. Here we are taking into account the fact that the dimension 4 operators which could possibly turn a slepton into a pair of leptons or a quark-antiquark pair are absent in this model.

{}Next, let us explain why it suffices to consider only operators of the type (\ref{higher}). The point is that all the other operators (allowed by the $SU(3)_c\otimes SU(2)_w\otimes U(1)_Y$ symmetry) which violate the baryon number by one unit and lepton number by (an odd) integer can be obtained by combinations of the following substitutions: $Q\leftrightarrow  D^* H_+$, $Q\leftrightarrow  U^* H_-$, $L\leftrightarrow  E^* H_+$, $L\leftrightarrow  N^* H_-$, 
$N\leftrightarrow  L^* H_-$, $N^*\leftrightarrow  L H_+$, as well as the obvious substitutions $H_+H_-\leftrightarrow 1$, $NN^*\leftrightarrow 1$, {\em etc}. None of these substitutions alter the discrete quantum numbers of a given operator, however. This implies that, if we show that the operators (\ref{higher}) are absent, then all the other dangerous operators are also forbidden.

{}It is now straightforward to show that proton is stable in our model. Indeed, the ${\widetilde{\cal L}}_3\otimes {\widetilde{\cal R}}_3$ discrete charges for the operators in (\ref{higher}) are given by $[k+1\mp m,k\mp m]$, respectively. Both of these charges cannot simultaneously be equal 0 modulo 3, so all such operators are forbidden by the discrete symmetries of the model. 

{}Here we should point out that the ${\widetilde{\cal L}}_3\otimes {\widetilde{\cal R}}_3$ discrete
symmetry is actually unnecessarily {\em strong} for just stabilizing proton. Indeed, consider the 
${\widetilde{\cal Y}}_3$ symmetry (which is a subgroup of ${\widetilde{\cal L}}_3\otimes {\widetilde{\cal R}}_3$ - see section III). The operators in (\ref{higher}) have +1 ${\widetilde{\cal Y}}_3$ charge regardless of the values of $m,k$. This implies that to stabilize proton it would suffice to consider the ${\widetilde{\cal Y}}_3$ symmetry instead of the larger ${\widetilde{\cal L}}_3\otimes {\widetilde{\cal R}}_3$ symmetry. As we already mentioned in the previous subsection, however, the former does {\em  not} forbid the dangerous dimension 5 operator $LLH_+H_+$ which is disastrous for the neutrino masses. In subsection E we will discuss a possibility of augmenting the ${\widetilde{\cal Y}}_3$ symmetry by a continuous flavor symmetry (broken on a ``distant'' brane) which may solve the problem with the dimension 5 operator $LLH_+H_+$ in an alternative way. At any rate, the full ${\widetilde{\cal L}}_3\otimes {\widetilde{\cal R}}_3$ discrete gauge symmetry solves both proton stability and neutrino mass problems in one shot.

\subsection{Additional ``Bonuses''}

{}In this subsection we will show that in the TSSM we have a possibility of explaining the mass hierarchy between the up and down quarks without appealing to large $\tan \beta$. Moreover, the $\mu$-term for the electroweak Higgs doublets $H_\pm$ as well as the $\mu^\prime$-term for the new states $F_\pm$ can also be generated.

{}Thus, let us consider the TSSM with the $SU(3)_c\otimes SU(2)_w\otimes U(1)_Y\otimes {\widetilde{\cal L}}_3\otimes {\widetilde{\cal R}}_3$ gauge symmetry. As we discussed in section III, the discrete gauge symmetries ${\widetilde{\cal L}}_3$ and ${\widetilde{\cal R}}_3$ can be viewed as remnants of the corresponding broken $U(1)_{\cal L}$ respectively $U(1)_{\cal R}$ gauge symmetries. Here we would like to take this a bit further and assume that the ${\widetilde{\cal L}}_3$ and/or ${\widetilde{\cal R}}_3$ discrete gauge symmetries indeed come from the corresponding continuous gauge symmetries, that is, $U(1)_{\cal L}$ and/or $U(1)_{\cal R}$ indeed exist in the Higgs phase with the corresponding massive gauge boson(s). 

{}Let us first consider the ${\widetilde{\cal L}}_3$ symmetry and the corresponding $U(1)_{\cal L}$ symmetry in this context.
Before the $U(1)_{\cal L}$ breaking we have, along with the chiral superfields $Q,D,U,L,E,N,H_\pm,F_\pm$, three additional fields neutral under $SU(3)_c\otimes SU(2)_w\otimes U(1)_Y$ (as well as ${\widetilde{\cal R}}_3$/$U(1)_{\cal R}$). These fields have the following charges under $U(1)_{\cal L}$:
\begin{equation}
 S: +3~,~~~S^\prime : +3~,~~~S^{\prime\prime}: -3~.
\end{equation}  
In the following we assume that the fields $S,S^\prime,S^{\prime\prime}$ are localized on the same set of $p$-branes as the chiral superfields $Q,D,U,L,E,H_\pm,F_\pm$. (The superfields $N_i$ are {\em not} localized on these branes - see below.)
Let us write down all the lowest dimensional couplings with and without the superfields $S,S^\prime,S^{\prime\prime}$ with the fields charges under $SU(3)_c\otimes SU(2)_w\otimes U(1)_Y$ allowed by all the gauge symmetries (for simplicity our notations here are symbolic, and we suppress the corresponding couplings, factors of $1/M_s$, and flavor indices):
\begin{eqnarray}
 && \left[(S+S^\prime) H_+H_-  +S^{\prime\prime} F_+ F_- + ~,\right.\nonumber\\
 && \left. 
 (S+S^\prime) L E H_- + L N H_+  
 +(S+S^\prime) Q D H_- + Q U H_+\right]_F ~,
\end{eqnarray}
where the actual linear combinations
of $S,S^\prime$ need not be the same in all three couplings involving these superfields. For definiteness let us assume that the $U(1)_{\cal L}$ breaking occurs in the unbroken supersymmetry limit. Then there is a flat direction along which the fields 
$S,S^\prime,S^{\prime\prime}$ can acquire non-zero VEVs\footnote{Here we are assuming that the mass term $mSS^{\prime\prime}+m^\prime S^\prime S^{\prime\prime}$ is absent, that is, these singlets are {\em moduli}. This is not an unreasonable assumption as in string models typically there is (embarrassing) proliferation of such states.}. The D-flatness condition requires that $|S|^2+|S^\prime|^2=|S^{\prime\prime}|^2$, and all the F-flatness conditions are satisfied as in the unbroken supersymmetry limit none of the fields (including $H_\pm$) that $S,S^\prime,S^{\prime\prime}$ couple to have non-zero VEVs. It then follows that at least one of the VEVs $S,S^\prime$ is of the same order of magnitude as $S^{\prime\prime}$. (In fact, generically, that is, without fine tuning, one could even assume that all three VEVs are in the same bulk range, but this will be unnecessary here.) Let us assume that (as a result of some non-trivial dynamics) the VEV of $S^{\prime\prime}$ is somewhat lower than $M_s$. Then the effective Yukawa couplings for the operators $L E H_-$ and $Q D H_-$ generated after the $U(1)_{\cal L}$ breaking will be suppressed by $M_{\cal L}/M_s$, where $M_{\cal L}$ is the $U(1)_{\cal L}$ breaking scale. This could explain the hierarchy between the up and down quark mass matrices
(as well as between the up quark and lepton mass matrices) {\em without} appealing to large $\tan \beta$. Also note that in the process of $U(1)_{\cal L}$ breaking the $\mu$-term for $H_\pm$ and the $\mu^\prime$-term for $F_\pm$ are automatically generated. 

{}It should now be clear why the second anomaly free $U(1)_{\cal L}$ charge assignment pointed out in footnote \ref{foot} is not appealing. There the charges of $H_+$ and $H_-$ 
(as well as of $F_+$ and $F_-$) are interchanged leading to reversing the hierarchy between the up and down quark mass matrices: with this second charge assignment the bottom quark would generically be heavier than the top quark.

{}Before we leave the subject of $U(1)_{\cal L}$ breaking, we would like to consider the more general case of breaking $U(1)_{\cal L}$ down to its ${\bf Z}_N$ subgroup ${\widetilde {\cal L}}_N$, where $N\not=3$. In this case all the fields except for the $S^\prime,S^{\prime\prime}$ have the same $U(1)_{\cal L}$ charges as before, but the charges of the superfields $S^\prime,S^{\prime\prime}$ now are $+N$ and $-N$, respectively. This means that we could generate the $\mu$-term and as well as the up-down hierarchy in such a model as the $S$ field can still acquire a non-zero VEV, but the $\mu^\prime$-term for the $F_\pm$ fields would be absent unless $N$ is a multiple of 3. However, if $N$ is not a multiple of three, $U(1)_{\cal L}$ is actually broken completely without a residual discrete symmetry.
The only way around this would be to assume that $N$ is a multiple of 3, and ${\widetilde {\cal L}}_N$ is broken down to ${\widetilde {\cal L}}_3$, in which case we are back to the ${\widetilde {\cal L}}_3$ discrete symmetry. (We could have the unbroken ${\widetilde {\cal L}}_N$ symmetry with $N\not=3$ at the expense of setting the VEV of $S$ to zero, but in that case the down quarks and leptons would be massless, and the $\mu$-term and the $\mu^\prime$-term would vanish as the corresponding operators would be forbidden by the ${\widetilde {\cal L}}_N$ symmetry.) Similar remarks apply to breaking $U(1)_{p{\cal L}+q{\cal R}}$
with $p\not=0$. This is the reason why we have been emphasizing the ${\widetilde {\cal L}}_3, {\widetilde {\cal X}}_3,{\widetilde {\cal Y}}_3$ discrete symmetries (and not their $N\not=3$ ${\bf Z}_N$ counterparts) in the previous discussions.

{}Note that the (possible) breaking of the $U(1)_{\cal R}$ gauge symmetry to its ${\widetilde {\cal R}}_3$ (or, more generally, ${\widetilde {\cal R}}_N$) discrete subgroup has no effect on the previous discussion of the $U(1)_{\cal L}$ breaking and its implications - due to the $U(1)_{\cal R}$ charge assignments, the fields that acquire VEVs to break $U(1)_{\cal R}$ couple very differently to the chiral superfields $Q,D,U,L,E,N,H_\pm,F_\pm$ compared with the $S,S^\prime,S^{\prime\prime}$ superfields discussed above.

{}Finally, we would like to discuss one other implication of breaking $U(1)_{\cal L}$ and/or $U(1)_{\cal R}$. Note that the right-handed neutrinos $N_i$ are charged under both of these symmetries. Now, the higher dimensional mechanism for generating small Dirac neutrino masses requires that the right-handed neutrinos propagate in the large ``bulk'' directions transverse to the $p$-branes on which the rest of the fields $Q,D,U,L,E,H_\pm,F_\pm$ are localized. This implies that $U(1)_{\cal L}$ as well as $U(1)_{\cal R}$ must be ``bulk'' gauge symmetries\footnote{This follows from the fact that by assumption the $N_i$ states can carry the corresponding flux away from the $p$-branes which makes it impossible to localize the $U(1)_{\cal L}$ and $U(1)_{\cal R}$ gauge bosons on these $p$-branes.}. Let us assume that the fields acquiring VEVs to break the $U(1)_{\cal L}$ and $U(1)_{\cal R}$ gauge symmetries are localized on the $p$-branes. This is actually necessary so that these gauge symmetries on the $p$-branes are reduced to their discrete remnants. Then the breaking of the ``bulk'' gauge symmetries is suppressed by the volume of the large transverse dimensions \cite{TeVphen}. Ignoring various numerical factors (as well as the fact that the compact direction(s) inside of the $p$-branes are somewhat large themselves), we can estimate the masses of the $U(1)_{\cal L}$ and $U(1)_{\cal R}$ gauge bosons to be of order of $M_s^2/M_P$, where $M_P$ is the four dimensional Planck scale. (Here we are assuming that the $U(1)_{\cal L}$ and $U(1)_{\cal R}$ breaking VEVs are roughly around $M_s$. In actuality these VEVs could be somewhat lower, but here we are only interested in a {\em very} rough estimate.) This implies that at sub-millimeter distances there are going to be gauge forces which can compete with gravity. The couplings of these ``bulk'' gauge bosons are very small (also suppressed by the volume of the transverse directions) - this is precisely why they would not be presently observable. However, these new forces can be several orders of magnitude stronger than gravity at sub-millimeter distances \cite{TeVphen}, and might be detectable in the upcoming measurements of the gravitational strength forces at such distances \cite{Price}.  
  
\subsection{Other Possibilities}

{}In this subsection we would like to briefly discuss a few possible variations of the solution to the proton decay and neutrino mass problems proposed in subsection C as some of the phenomenological implications of these ``alternative'' solutions might be different.

\begin{center}
 {\em Higher Discrete Symmetries}
\end{center}

{}One of the most obvious other possibilities is to gauge higher discrete symmetries instead of ${\bf Z}_3$ (or, more, precisely, ${\bf Z}_3\otimes {\bf Z}_3$ as in subsection C). From the previous subsection it should be clear that we are stuck with ${\widetilde{\cal L}}_3$ (as far as the ${\widetilde{\cal L}}_N$ part is concerned). Also, as we pointed out in subsection A, the ${\widetilde{\cal R}}_N$ symmetry by itself cannot do the job. We can therefore consider ${\widetilde{\cal L}}_3\otimes {\widetilde{\cal R}}_N$ symmetries. It is not difficult to see that if $N$ is a multiple of 3 this symmetry forbids all the operators possibly leading to proton decay. As to the allowed dimension 3 and 4 operators, they are the same in this case as for the ${\widetilde{\cal L}}_3\otimes {\widetilde{\cal R}}_3$ symmetry.

{}On the other hand, if we gauge ${\widetilde{\cal L}}_3\otimes {\widetilde{\cal R}}_N$ for $N$ not a multiple of 3, some of the operators leading to proton decay are no longer absent. In this case one has to carefully check whether the proton decay rate is sufficiently suppressed. We will not perform such an analysis here for this case as it is similar to that discussed for the next possibility.     

\begin{center}
 {\em The ${\widetilde{\cal X}}_3$ Symmetry}
\end{center}

{}We can try to consider the ${\bf Z}_3$ subgroup of ${\widetilde{\cal L}}_3\otimes {\widetilde{\cal R}}_3$ which in section III we denoted by ${\widetilde{\cal X}}_3$. This discrete symmetry forbids the dangerous dimension 5 operator $LLH_+H_+$. However, it does {\em not} forbid all higher dimensional operators possibly leading to proton decay. At first it might seem that the lowest dimension operator of this type is $QQQLLH_+$ (or those obtained from this one by substitutions described in subsection C). However, this operator contains an even number of $L$ insertions, and dimension 4 operators which could turn a slepton into a pair of leptons  
or a quark-antiquark pair are forbidden by the ${\widetilde{\cal X}}_3$
symmetry. So the lowest dimension operator (allowed by the
${\widetilde{\cal X}}_3$ symmetry) which is dangerous for proton decay
contains five $L$ insertions\footnote{Here the following remark is in
order. We can trade some of the $L$ insertions for the $N^*$ insertions via
the substitution $LH_+\rightarrow N^*$. This way we can obtain, say, the
operator $QQQLN^{*4}$ which naively might appear to be less suppressed than
the $QQQL^5 H_+^4$ operator for the former does not have extra $H_+$
insertions. However, since the right-handed neutrinos are ``bulk'' fields,
their couplings to the fields localized on the $p$-branes such as charged
quarks and leptons are further suppressed by the volume of the large
transverse dimensions. On the other hand, there is an enhancement factor
due to the large number of the right-handed neutrino KK modes in the
transverse directions. The net result of these two competing effects is
that each $N^*$ (or $N$) insertion is suppressed by an additional factor
(or, more generally, power) of $(m_p/M_s)$ ($m_p$ is the proton mass) compared with other insertions such as $L$. This implies that we need not worry about the proton decay channels with one or more right-handed neutrinos in the final state.}: $QQQL^5H_+^4$. If we ignore the fact that the VEV of $H_+$ is somewhat smaller than $M_s$ (that is, if we set $\langle H_+\rangle\sim M_s$, we can get a rough estimate for the proton decay rate: $\Gamma_p/m_p\sim (m_p/M_s)^{10}$.
(Here $m_p$ is the proton mass.) Here we should point out that this estimate is {\em very} rough not only for the reason that we have not taken into account the $\langle H_+\rangle/M_s$ suppression factors, but also for other reasons such as additional loop suppression factors (including extra powers of the gauge couplings and various numerical factors) as well as the phase space suppression factors. Nonetheless, let us estimate the conservative lower bound on the string scale we obtain this way. Let us require that $\Gamma_p/m_p{\ \lower-1.2pt\vbox{\hbox{\rlap{$<$}\lower5pt\vbox{\hbox{$\sim$}}}}\ } 10^{-65}$, which implies that the proton lifetime $\tau_p{\ \lower-1.2pt\vbox{\hbox{\rlap{$>$}\lower5pt\vbox{\hbox{$\sim$}}}}\ } 2\times 10^{33}~{\mbox{years}}$. Then we get the following lower bound on the string scale: $M_s
{\ \lower-1.2pt\vbox{\hbox{\rlap{$>$}\lower5pt\vbox{\hbox{$\sim$}}}}\ } 3\times10^3~{\mbox{TeV}}$. This estimate, however, is too conservative - we can lower this bound by including the $\langle H_+\rangle/M_s$ suppression factors. The new estimate for the decay rate due to the operator $QQQL^5H_+^4$ then is $\Gamma_p/m_p\sim (m_p/M_s)^{10}\times (M_{ew}/M_s)^8$, where we will take $M_{ew}\sim 250~{\mbox{GeV}}$. 
Then the lower bound is brought down to $M_s
{\ \lower-1.2pt\vbox{\hbox{\rlap{$>$}\lower5pt\vbox{\hbox{$\sim$}}}}\ } 50~{\mbox{TeV}}$. However, the operator $QQQL^5H_+^4$ is not the only dangerous operator we must consider. There are other {\em a priori} less suppressed operators such as $U^* U^* D^* L^5 H_+^3$ and
$U^* U^* Q L^5 H_+^2$. The rough estimate for these two operators gives the lower bounds
$M_s {\ \lower-1.2pt\vbox{\hbox{\rlap{$>$}\lower5pt\vbox{\hbox{$\sim$}}}}\ } 90~{\mbox{TeV}}$
respectively $M_s {\ \lower-1.2pt\vbox{\hbox{\rlap{$>$}\lower5pt\vbox{\hbox{$\sim$}}}}\ } 210~{\mbox{TeV}}$. Here we should point out that in both of these operators two of the $L$ insertions must contract their $SU(2)_w$ indices with each other meaning that  they are antisymmetrized. This might lead to an additional (``flavor'') suppression factor. At any rate, our estimates here are very rough, and are only intended for illustrative purposes. It is {\em a priori} possible that in such a model the proton decay rate might be suppressed to an acceptable level for $M_s$ between 10 and 100 TeV. Here we should mention that the dimension 3 and 4 operators
allowed by the ${\widetilde{\cal X}}_3$ symmetry are the same as in the ${\widetilde{\cal L}}_3\otimes {\widetilde{\cal R}}_3$ case.     

\begin{center}
 {\em Discrete Plus $U(1)$ Gauge Symmetries}
\end{center}

{}Another possibility is to take a discrete gauge symmetry that cannot suppress proton decay to an acceptable level on its own and augment it with a continuous gauge symmetry. The latter must be a ``bulk'' gauge symmetry which is broken on a ``distant'' brane along the lines of \cite{TeVphen}. As an example of such a scenario consider gauging\footnote{If we consider gauging just ${\widetilde {\cal L}}_3$, the proton decay rate would be unacceptably high for $M_s$ in the TeV range. Thus, already the operator $QQQLLLH_+H_+$ (which is allowed by the ${\widetilde {\cal L}}_3$ symmetry) would be problematic: to suppress the proton decay rate adequately we would have to assume that $M_s{\ \lower-1.2pt\vbox{\hbox{\rlap{$>$}\lower5pt\vbox{\hbox{$\sim$}}}}\ } 10^4~{\mbox{TeV}}$.} ${\widetilde{\cal L}}_3\otimes U(1)_{\cal Z}$, where ${\cal Z}=Y-{\cal R}$ is the $B-L$ (baryon minus lepton number) generator discussed in section III. Let the corresponding $U(1)_{\cal Z}$ be a ``bulk'' gauge symmetry which is broken by non-zero VEVs of some fields localized on a ``distant'' brane. Then, provided that proton decay is absent in the limit where the ${\widetilde{\cal L}}_3\otimes U(1)_{\cal Z}$ symmetry is unbroken, the proton decay rate induced by the $U(1)_{\cal Z}$ breaking on the ``distant'' brane can be sufficiently suppressed (at least by some power of the volume of the large transverse directions). To show that the unbroken ${\widetilde{\cal L}}_3\otimes U(1)_{\cal Z}$ symmetry stabilizes proton, note that ${\widetilde{\cal L}}_3$ by itself forbids all operators (\ref{higher}) except for those violating the lepton number conservation by multiples of 3. (In particular, the ${\widetilde{\cal L}}_3$ symmetry forbids the dangerous dimension 5 operator $LLH_+H_+$.) Since these operators violate baryon number conservation by one unit, the operators allowed by ${\widetilde{\cal L}}_3$ always violate the   
$U(1)_{\cal Z}$ charge conservation, and therefore are all forbidden. As pointed out in \cite{TeVphen}, gauging $U(1)_{\cal Z}$ in the ``bulk'' has interesting experimental consequences. Thus, at sub-millimeter distances one expects an isotope dependent new force - the baryon minus lepton number of a neutral atom is the number of neutrons in its nucleus.

\begin{center}
 {\em Discrete Plus Non-Abelian ``Flavor'' Gauge Symmetries}
\end{center}

{}Finally, we would like to consider a possibility of combining a discrete gauge symmetry with a non-Abelian gauge symmetry. The latter (or some part of it) would have to be gauged in the ``bulk'' and broken on a ``distant'' brane. As an example of such a scenario consider gauging
${\widetilde{\cal Y}}_3$ together with some non-Abelian symmetry which we
will specify in a moment. First note, however, that, as we pointed out in
subsection C, the ${\widetilde{\cal Y}}_3$ discrete gauge symmetry by
itself stabilizes proton. The only trouble with this symmetry is that it
does not forbid the dimension 5 operator $LLH_+H_+$. Gauging, say,
${\widetilde{\cal L}}_3$ on top of ${\widetilde{\cal Y}}_3$ (which would
solve the problem) would be equivalent to gauging ${\widetilde{\cal
L}}_3\otimes {\widetilde{\cal R}}_3$ which we already considered in
subsection C. We could gauge, say, $U(1)_{\cal F}$ in the ``bulk'' as in
the previous scenario. This would also solve the problem, and this
possibility falls under the general category of solutions involving
discrete plus $U(1)$ gauge symmetries. However, here we can imagine gauging
a non-Abelian symmetry such as a flavor symmetry which could forbid the
operator $LLH_+H_+$. Thus, consider the following possibility briefly
discussed in \cite{neutrino}\footnote{I am grateful to Gia Dvali for a
discussion on this point.}. Suppose that in the lepton sector the flavor
gauge symmetry is $U(2)_L\otimes U(2)_R$. Here $U(2)_L$ can be localized on
the same set of $p$-branes as quarks and leptons, whereas $U(2)_R$ can be a
``bulk'' gauge symmetry. The $U(2)_L\otimes U(2)_R$ flavor gauge symmetry
must somehow be broken. Consider breaking this symmetry by fields in the
bifundamental representation which are localized on the $p$-branes, as well
as some other fields localized on a ``distant'' brane. Then the operator $LLH_+H_+$ can be adequately suppressed - note that if $U(2)_L\otimes U(2)_R$ is only broken by bifundamentals, this operator is forbidden altogether, while the charged lepton masses still can arise. This implies that one can arrange breaking the flavor symmetry completely (meaning, without any residual massless gauge bosons), while generating non-degenerate mass matrices in the charged lepton sector, and at the same time suppressing the Majorana masses adequately. An interesting implication of such a scenario is that the neutrino magnetic moment operator is {\em not} suppressed even though the Majorana masses are \cite{neutrino}. That is, generically the neutrino magnetic moment in this scenario can be large (as it is suppressed only by $M_s$) compared with the usual ``see-saw'' scenario (the latter being considered in the $M_s\sim M_{\small{GUT}}$ context).  

{}Here we should point out that the question of gauging non-Abelian flavor symmetries in the TeV scale scenario has been addressed in \cite{flavor1,flavor}. There appear to be some open questions related to the problem of suppressing flavor changing neutral currents (FCNCs) \cite{flavor}.
This, in a way, seems to be related to the fact that the flavor gauge symmetries must be broken at the end of the day. For this reason it is not completely clear whether the above scenario will pass all the tests as far FCNCs are concerned - it is not unlikely that the flavor symmetries in the lepton and quark sectors are related. At any rate, the FCNC problem is beyond the scope of this paper, so we will leave the question of viability of this scenario for the future investigations.  

\subsection{Signatures of High {\em vs.} Low $M_s$ Scenarios}

{}In this subsection we would like to address the following question: do we expect any signatures for the low $M_s$ scenarios that would be very distinct from those for the high $M_s$ scenarios? More concretely, can we expect that the upcoming collider experiments will be able to distinguish between these scenarios even {\em without} a direct production of heavy Kaluza-Klein or string states? Here we would like to point out that the answer to these questions might be positive.

{}Let us first consider the high $M_s$ scenarios where $M_s$ is of order $M_{\small{GUT}}$. In fact, as we pointed out in \cite{zura}, this is possible even in the model we have been discussing in this paper: if the mass scale of the new states $F_\pm$ is high\footnote{Here we should point out the following. If one repeats the analyses of subsection D, it is not difficult to see that in our model obtaining a large $\mu^\prime$-term for the $F_\pm$ states while keeping a low $\mu$-term for the $H_\pm$ states as well as having the desired Yukawa couplings for the quarks and leptons with the Higgs doublets in our model would require an unnatural conspiracy between the couplings of the $S,S^\prime$ states to $H_\pm$ as well as the VEVs of the former. From this viewpoint in our model it seems more natural to have low $M_s$ rather than high $M_s$.}, that is, of order $M_{\small{GUT}}$ or around that scale, then the unification of couplings can occur at a high $M_s$ scale. This is clear for the reason that in this case the usual logarithmic running of gauge couplings in this model would be the same as in the MSSM, and the power-like running above the KK threshold would simply lower the unification scale by some factor which can be of order one. In fact, we could even consider a model with no contribution from the heavy KK states to the gauge coupling running - as far as the low energy measurements of the gauge couplings are concerned this would be immaterial. The presence of the heavy $F_\pm$ states might still have some relevance as they may be responsible for the discrete anomaly cancellation as we discussed in section III. Nonetheless, since the scale of these new states has been pushed up so high, it is {\em a priori} possible to have completely different states that do the same job of the discrete anomaly  cancellation (examples of which can be found in \cite{IR}) without spoiling any of the other features of the theory, or contradicting the low energy data. To summarize, once we push the string scale up to $M_{\small{GUT}}$ the main motivations for considering our model disappear - this is not surprising as most of the low energy physics ``decouples'' from the high energy physics. Can we then make any definitive conclusions about whether the string scale is high or low by examining the low energy data?

{}One process that constraints {\em both} high and low energy physics is proton decay. As we already mentioned in subsection A, certain dimension 5 operators such as $QQQL$ are dangerous for proton stability {\em even} if $M_s$ is as high as $M_{\small{GUT}}$. That is, proton stability constraint is sensitive to high energy physics, and in this sense the latter does not ``decouple''. On the other hand, suppose we are considering a high $M_s$ scenario. Then generically we must allow the $LLH_+H_+$ operator to be present to generate the desirable neutrino masses via some kind of ``see-saw'' mechanism. As to the discrete gauge symmetries that we considered in section III, the only one that can do the job of suppressing the dangerous baryon plus lepton number violating dimension 5 operators {\em without} at the same time forbidding the $LLH_+H_+$ operator is the ${\widetilde {\cal Y}}_3$ symmetry. This was actually already observed in \cite{IR}. The implication of this fact could be quite interesting: the point is that the ${\widetilde {\cal Y}}_3$ symmetry allows all dimension 3 and 4 lepton number violating operators to be present (but it forbids the dimension 4 baryon number violating operator). That is, the upcoming collider experiments are likely to see lepton number violating processes via the slepton channel provided that $M_s$ is high. 

{}Let us compare this prediction to those expected in the low $M_s$ scenarios. Here we need to suppress proton decay even more, but the situation with the neutrino masses is reversed. The $LLH_+H_+$ operator that we desired to keep in the high $M_s$ scenarios would now be disastrous, so we have to forbid this operator. All of the solutions we discussed in the previous subsections except for one forbid all the dimension 3 and 4 lepton number violating operators.           
The only exception is the case where we gauge the ${\widetilde {\cal Y}}_3$ symmetry together with some non-Abelian flavor gauge symmetry. In that case we can still have lepton number violating dimension 3 and 4 operators. However, some of the operators allowed without the additional flavor symmetry are now suppressed. So even in this case the collider signatures of the low $M_s$ scenarios should be quite different. Another interesting point is that in the low $M_s$ scenario with the ${\widetilde {\cal Y}}_3$ discrete symmetry and $U(2)_L\otimes U(2)_R$ flavor gauge symmetry the neutrino magnetic moments are generically expected to be much larger than in the high $M_s$ scenarios. Also note that one of the key differences is the mass scale of the new $F_\pm$ states which are expected to be present in the low $M_s$ scenarios. In contrast, in the high $M_s$ scenarios these states should either be absent altogether, or must be very heavy (or else the usual MSSM unification prediction would be ruined). To summarize, the upcoming collider experiments (as well as other types of experiments) might be able to indirectly probe how high $M_s$ is without actually producing heavy KK or string states. This is especially encouraging as the nearest future collider experiments will not be able to directly probe the KK or string modes if they are heavier than a few TeV. 

\section{Brane World Embedding}

{}In the previous sections we discussed the TSSM where, as was shown in \cite{zura}, the gauge coupling unification occurs in the TeV range. One of the key ingredients of the TSSM is the presence of new (compared with the MSSM spectrum) states $F_\pm$ which are crucial for the fact that the unification in the TSSM is as precise (at one loop) as in the MSSM. In this paper we have shown that introduction of these states allows to gauge anomaly free discrete as well as continuous symmetries which stabilize proton and are also important for successful generation of small neutrino masses in the TSSM. 

{}In this section  we would like to discuss possible embeddings of the TSSM in the brane world framework. Some of these points were discussed in \cite{zura}. Here we will briefly review the general discussion of \cite{zura}, and then we will focus on more concrete issues such as how to obtain in the brane world context the discrete (as well as continuous) gauge symmetries we have used for stabilizing proton in the TSSM.

{}Note that at present it is not known how to explicitly construct a string vacuum with all the desirable features of the TSSM. This, however, by no means implies that such a vacuum does not exist in string theory. Rather, it might be related to the present lack of the necessary technology for such model building. In fact, as was already pointed out in \cite{zura}, such a vacuum is not expected to have a perturbative description which makes its explicit construction a non-trivial task. Moreover, many of the consistent string vacua have not yet been understood. Our discussion in this section is aimed at hopefully obtaining hints which might be useful in narrowing down an {\em a priori} vast choice of possibilities by imposing various phenomenological  constraints.

\subsection{Generalized Voisin-Borcea Orbifolds}      

{}One of the simplest ways of obtaining models with some of the features discussed in the previous sections is via Type I (or Type I$^\prime$)\footnote{For recent developments in four dimensional Type I (Type I$^\prime$) compactifications/orientifolds, see, {\em e.g.}, \cite{typeI,KST,3gen,3gen1}.} compactifications on generalized
Voisin-Borcea orbifolds. Here we would like to review some facts about these Calabi-Yau three-folds. Let ${\cal W}_2$ be a K3 surface (which is not necessarily an 
orbifold) which admits a ${\bf Z}_N$ action $J$ such that $J\Omega=\alpha^{-1} \Omega$,
where $\Omega$ is the holomorphic
two-form $dz_1\wedge dz_2$ on ${\cal W}_2$, and $\alpha=\exp(2\pi i/N)$. Consider the following quotient:
\begin{equation}
 {\cal Y}_3= (T^2\otimes {\cal W}_2)/Y~,
\end{equation}
where $Y=\{S^k|k=0,\dots,N-1\}\approx {\bf Z}_N$, and $S$ acts as $g z_0=\alpha z_0$ on $T^2$ 
($z_0$ being a complex coordinate on $T^2$), and as $J$ on ${\cal W}_2$. This
quotient is a Calabi-Yau three-fold with $SU(3)$ holonomy which is elliptically 
fibered over the base ${\cal B}_2={\cal W}_2/B$ 
($B=\{J^k|k=0,\dots,N-1\}\approx{\bf Z}_N$) provided that $g$ acts crystallographically on $T^2$. This implies that such Calabi-Yau quotients exist only for $N=2,3,4,6$. In the $N=2$ case we have the original Voisin-Borcea orbifolds \cite{Voisin}. 

{}Next, consider Type I compactification on a generalized Voisin-Borcea orbifold ${\cal Y}_3$. Here one needs to specify the gauge bundle embedded in the D9-brane gauge group. For certain choices of the gauge bundle as well as the base ${\cal B}_2$, we can also have D5-branes. Here we are going to be interested in D5-branes wrapped on the fibre $T^2$. The low energy four dimensional gauge theory living in the world-volume of these D5-branes has ${\cal N}=1$ supersymmetry. We also have KK states corresponding to compactifying the original six dimensional D5-brane world-volume theory on $T^2$. More precisely, we can think about these KK modes as follows. First consider Type I compactified on K3. This six dimensional theory has ${\cal N}=1$ supersymmetry. Next, consider D5-branes whose transverse directions correspond to K3. That is, these D5-branes fill the six dimensional Minkowski space ${\bf R}^{5,1}$. Let us now further compactify two of the directions in ${\bf R}^{5,1}$ on a two-torus $T^2$. The corresponding four dimensional theory has ${\cal N}=2$ supersymmetry. Thus, the low energy effective theory of the D5-brane world-volume theory is an ${\cal N}=2$ gauge theory in four dimensions. The corresponding KK excitations also have ${\cal N}=2$ supersymmetry from the four dimensional viewpoint. Now let us orbifold this theory by the ${\bf Z}_N$ orbifold group $Y$ whose generator acts as a ${\bf Z}_N$ rotation $z_0\rightarrow \alpha z_0$ on $T^2$, and as  $J$ on K3. The latter breaks the $SU(2)_R$ R-parity group of the ${\cal N}=2$ gauge theory to $U(1)_R$. Correlated together with the rotation on $T^2$, it produces an ${\cal N}=1$ supersymmetric gauge theory in four dimensions plus the KK excitations which still come in ${\cal N}=2$ supermultiplets. The last statement follows from the fact that the ${\bf Z}_N$ action on the heavy KK modes from $T^2$ acts by rotating the KK states with different momenta in the $z_0$ direction into each other, and the resulting states are given by linear combinations of the former invariant under ${\bf Z}_N$ rotations. This, in particular, implies that the number of heavy KK modes is reduced by $N$ compared with the parent ${\cal N}=2$ theory. 

{}Note that perturbatively, when considering Type I compactifications on orbifolds, the orbifold group action on the Chan-Paton factors is typically fixed by the tadpole cancellation conditions. More precisely, suppose we view a given Type I compactification as a Type IIB orientifold. Then generically the action of the orientifold on the Chan-Paton factors is determined (or at least severely constrained) by the one-loop tadpole cancellation conditions. On the other hand, as pointed out in \cite{zura}, to obtain the desired KK spectrum in our model we must assume that
the ${\bf Z}_N$ orbifold action on the gauge quantum numbers is trivial. In \cite{zura} this was used to argue that if there is an orientifold embedding for our model it should be non-perturbative. This is further supported by the observations presented in the next subsection. 

\subsection{Twisted Sectors and Discrete Gauge Symmetries}

{}As we mentioned in section II, some of the chiral matter fields, namely, the states $Q_i,D_i,U_i,L_i,E_i$ in the light spectrum 
of the TSSM are localized in $3+1$ dimensions even though the gauge bosons of $SU(3)_c\otimes SU(2)_w\otimes U(1)_Y$ as well as the states $H_\pm$ and $F_\pm$ arise upon compactifying the $5+1$ dimensional gauge theory down to four dimensions. (The compactification space here is $T^2/{\bf Z}_N$.)  
That is, the above matter fields have no KK counterparts corresponding to the two compact directions inside of the D5-branes, whereas the gauge bosons plus the $H_\pm,F_\pm$ states do. Here we can ask whether this can occur in string theory. The answer to this question is positive in the following sense. Since we are considering orbifold compactifications, the corresponding quotients will have a set of points fixed under the action of the orbifold group. For instance, in the case of the generalized Voisin-Borcea orbifolds the ${\bf Z}_N$ twist $S$ (see the previous subsection) can act with fixed points. At each fixed point there is a collapsed two-sphere ${\bf P}^1$. If the world-sheet description is adequate D-branes wrapped on such ${\bf P}^1$'s do not give rise to non-perturbative states (which is due to the B-flux stuck inside of the ${\bf P}^1$'s \cite{aspin}).      
However, as was argued in \cite{KST}, for certain compactifications we must turn off the B-flux, and wrapped D-branes do give rise to non-perturbative ``twisted'' sector states. Note that in the case of the generalized Voisin-Borcea orbifolds in the present context these states would live in the non-compact ${\bf R}^{3,1}$ part of the D5-brane world-volume, but they are localized at points on the fibre $T^2$ which the D5-branes wrap. This implies that such light twisted sector states do not have KK counterparts corresponding to the fibre $T^2$. Thus, to obtain the additional chiral sectors with the lepton and quark quantum numbers we can consider certain {\em non-perturbative} orientifolds with twisted sectors. As explained in detail in \cite{KST}, such sectors do not possess world-sheet description. One way to understand such states is to consider the map \cite{sen} of (the T-dual of) the corresponding Type I vacuum to F-theory \cite{vafa}. (Such a map for the Voisin-Borcea orbifolds with $N=2$ was discussed in detail in \cite{KST}. For the cases with $N\not=2$ such a map can be more non-trivial.) 

{}Next, we would like to discuss the possible origins of the discrete gauge symmetries we have introduced in the TSSM. Such discrete gauge symmetries can be present in the orbifold compactifications, and are related to the orbifold group itself. For definiteness let us focus on the discrete gauge symmetries discussed in subsection C of section IV, that is, the ${\widetilde{\cal L}}_3\otimes {\widetilde{\cal R}}_3$ discrete gauge symmetry which is a ${\bf Z}_3\otimes {\bf Z}_3$ symmetry. Let us first discuss the possible origin of the ${\widetilde{\cal L}}_3$ 
discrete gauge symmetry. Note that the states $H_\pm$ and $F_\pm$ carry no charges under this symmetry. They, at the same time, have KK counterparts corresponding to the two compact directions inside of the D5-branes. Thus, it is natural to assume that these states come from the ${\bf Z}_N$ untwisted sectors, where $N=3$ or $N=6$ (here by ${\bf Z}_N$ we mean the orbifold group in the generalized Voisin-Borcea orbifold construction). In fact, for now let us consider the case $N=3$. Then the chiral matter fields $Q_i,D_i,U_i,L_i,E_i$ should come from the ${\bf Z}_3$ twisted sectors. Here two points should be stressed. First, the number of generations, which is three, is related to the number of fixed points of the ${\bf Z}_3$ twist. Generally, this number can be three or a multiple of three depending on details. In particular, the fibre $T^2$ part contributes a factor of three into the number of fixed points. If the corresponding factor from the base is one, we have three fixed points altogether, and this would give rise to three generations. (We will discuss the case where the base contribution does not equal one in a moment.) Second, we expect all of the fields $Q_i,D_i,U_i,L_i,E_i$ to carry non-zero discrete ${\bf Z}_3$ charges which we would like to identify with the ${\widetilde{\cal L}}_3$ charges. However, the fields $Q_i,D_i,U_i$ are neutral under ${\widetilde{\cal L}}_3$. Recall from appendix A, however, that the ${\bf Z}_3$ charges are only defined up to shifts (\ref{shift}) which precisely in the ${\bf Z}_3$ case do not affect the $H_\pm,F_\pm$ charge assignments but do affect those for $Q_i,D_i,U_i$ (the $L_i,E_i$ charges are not affected either). After such a shift all of the states $Q_i,D_i,U_i,L_i,E_i$ carry non-zero ${\bf Z}_3$ charges, and can be assumed to arise in the ${\bf Z}_3$ twisted sectors. To summarize, we propose here that the ${\widetilde{\cal L}}_3$ discrete gauge symmetry in our model can be identified with the ${\bf Z}_3$ symmetry arising due to orbifolding.       

{}Next, let us try to identify the origin of the second discrete gauge symmetry, that is, ${\widetilde{\cal R}}_3$. To have such a ${\bf Z}_3$ symmetry we need another orbifold action.
This can be arranged as follows. Let the K3 surface ${\cal W}_2$ mentioned above be given by the following quotient:
\begin{equation}
 {\cal W}_2 ={\widetilde {\cal W}}_2/X~,
\end{equation}
where $X=\{1,T,T^2\}\approx{\bf Z}_3^\prime$ is the orbifold group, and we are using prime to distinguish it from the ${\bf Z}_3$ orbifold generated by $S$. The surface ${\widetilde {\cal W}}_2$ can either be a four-torus or a different K3 surface. Note that the generator $T$ must leave the holomorphic two-form ${\widetilde \Omega}$ on ${\widetilde {\cal W}}_2$ invariant.

{}Next, we identify the ${\widetilde{\cal R}}_3$ charges with the ${\bf Z}_3^\prime$ discrete quantum numbers. Note that $H_\pm$ and $F_\pm$ carry non-zero charges under ${\widetilde{\cal R}}_3$. This implies that they come from ${\bf Z}_3^\prime$ twisted sectors, and must themselves be non-perturbative states. Below we show which twisted sector each of the states comes from (the ${\widetilde{\cal L}}_3$ charge, which has been shifted according to (\ref{shift}), is given by the power of $S$, whereas the ${\widetilde{\cal R}}_3$ charge is given by 
the power of $T$):
\begin{eqnarray}
 &&{\mbox{untwisted}}:~~~{\mbox{gauge bosons}},\nonumber\\
 &&S-{\mbox{twisted}}:~~~Q_i,L_i~,\nonumber\\
 &&T-{\mbox{twisted}}:~~~H_+,F_-~,\nonumber\\
 &&T^{-1}-{\mbox{twisted}}:~~~H_-,F_+~,\nonumber\\
 &&S^{-1}T-{\mbox{twisted}}:~~~D_i,E_i~,\nonumber\\
 &&S^{-1}T^{-1}-{\mbox{twisted}}:~~~U_i,N_i~,\nonumber
\end{eqnarray}
and there are no states coming from the twisted sectors that are not shown. There also exist the second solution corresponding to shifting the original ${\widetilde{\cal L}}_3$ charges according to (\ref{shift}) except in the direction opposite to that of the previous solution. This solution reads:
\begin{eqnarray}
 &&{\mbox{untwisted}}:~~~{\mbox{gauge bosons}},\nonumber\\
 &&S-{\mbox{twisted}}:~~~L_i~,\nonumber\\
 &&S^{-1}-{\mbox{twisted}}:~~~Q_i~,\nonumber\\
 &&T-{\mbox{twisted}}:~~~H_+,F_-~,\nonumber\\
 &&T^{-1}-{\mbox{twisted}}:~~~H_-,F_+~,\nonumber\\
 &&ST-{\mbox{twisted}}:~~~D_i~,\nonumber\\
 &&S^{-1}T^{-1}-{\mbox{twisted}}:~~~N_i~,\nonumber\\
 &&ST^{-1}-{\mbox{twisted}}:~~~U_i~,\nonumber\\
 &&S^{-1}T-{\mbox{twisted}}:~~~E_i~.\nonumber
\end{eqnarray}
Without a concrete model, however, it is unclear which particular solution can be realized in a given situation.

{}Here the following remark is in order. Consider such a non-perturbative compactification on a generalized Voisin-Borcea orbifold with $N=3$. Geometrically it is clear that all three chiral families will have identical Yukawa couplings with the electroweak Higgs doublets. This is not very appealing phenomenologically as all three generations are going to be top-like (or light) after the electroweak breaking. However, as was pointed out in \cite{zura}, one can discriminate
\cite{INQ} between the three fixed points on the fibre $T^2$ by turning on non-trivial Wilson lines which act trivially on the $SU(3)_c\otimes SU(2)_w\otimes U(1)_Y$ quantum numbers (which is required to maintain the unification prediction - see \cite{zura} for details), but can act non-trivially on some other gauge quantum numbers (corresponding, for instance, to hidden/horizontal 
gauge symmetries). This way different fixed points can give rise to chiral generations with different couplings to the electroweak Higgs doublets. Another possibility could be to break the ``accidental'' $SU(3)$ ``flavor'' symmetry between the three generations dynamically. An example of such a dynamical mechanism was given in \cite{fla}.

{}The above point about the degeneracy of Yukawa couplings is ameliorated if we consider ${\bf Z}_N$ orbifolds with $N=6$. Note that the desirable ${\bf Z}_3$ symmetry corresponding to 
${\widetilde{\cal L}}_3$ is not affected by the extra ${\bf Z}_2$ twist as ${\bf Z}_6\approx{\bf Z}_3\otimes {\bf Z}_2$. That is, there might exist an additional ${\bf Z}_2$ symmetry (which we have not utilized). Thus, we could consider a ${\bf Z}_6\otimes {\bf Z}_3^\prime$ or even ${\bf Z}_6\otimes {\bf Z}_6^\prime$ orbifold. Note that the number of fixed points on the fibre $T^2$ is no longer 3, so one might wonder whether we could still obtain three generations. Examples of ${\bf Z}_6$ orbifolds with three chiral families are known in the literature, see, {\em e.g.}, \cite{thr}. In such cases three generations are no longer degenerate, so one top-like generation can in principle arise without any additional complications.

{}Finally, we would like to comment on the right-handed neutrinos $N_i$ in 
the above picture. They are neutral under the 
$SU(3)_c\otimes SU(2)_w\otimes U(1)_Y$ gauge group 
(whose gauge bosons are localized on D5-branes), 
so they need {\em not} be localized on the D5-branes. 
Thus, they can have KK modes in the directions transverse 
to the D5-branes (note that at least two of these directions are 
supposed to be large to correctly reproduce the four 
dimensional Planck scale). Thus, the right-handed 
neutrinos $N_i$ can, for instance, come from the 
corresponding twisted {\em closed} string sector. 
However, it is not difficult to see that they can only either have 
KK modes in two directions transverse to the D5-branes or 
no KK modes at all (in the transverse directions - they can in principle 
have KK modes in the directions inside of the D5-branes).
This follows from the fact that to have ${\cal N}=1$ supersymmetry 
the corresponding twist must have a set of fixed points of real 
dimension 0 or 2. In the former case the fields $N_i$ are localized 
in $3+1$ dimensions and have no KK modes. In the latter 
case they can have KK modes in the two directions inside 
of the D5-branes and be localized at a point in the transverse 
directions, or have KK modes in the two directions transverse 
to the D5-branes and be localized at a point in the other four 
compact directions. Only the last case is phenomenologically 
acceptable if we would like to generate small Dirac neutrino 
masses via the mechanism of \cite{neutrino}. Moreover, to obtain 
the correct values for the neutrino masses, we must assume 
that there are only two large directions transverse to the D5-branes, 
and the fields $N_i$ have KK modes precisely in these directions. 
Thus, we cannot have four large transverse directions in this picture. 
It was pointed out in \cite{neutrino} that in the case of only two large 
transverse directions there might be some cosmological problems 
with the TeV-scale brane world scenario as the neutrino masses 
are of the same order of magnitude as the lowest KK masses, and 
the mixing between the two would imply that too many KK modes 
would be thermalized during the big-bang nucleosynthesis 
which would essentially jeopardize the latter. This argument, 
however, is based on the assumption that $M_s\sim 1~{\mbox{TeV}}$, 
and none of the other dimensions are large. In our case we have two 
relatively large dimensions inside of the D5-branes, and, as pointed 
out in \cite{zura}, the successful gauge coupling unification essentially 
requires that $M_s$ be in the $10-100~{\mbox{TeV}}$ range. Then it is 
not difficult to check that the neutrino masses are actually a few orders 
of magnitude smaller than the lowest KK masses, so that the mixing 
between the two is no longer large. This does not completely 
guarantee that thermalization of the KK modes is not dangerous, 
but depending on the details one could hope that the corresponding 
rate could be at least marginally acceptable. Note that if we could 
have more than two large transverse dimensions this problem 
(as well as other cosmological problems in the TeV-scale brane world scenario) 
would be ameliorated. It would therefore be interesting to see if one 
could accommodate the above picture with the right-handed neutrinos 
having KK modes in more than two transverse dimensions. We 
should note, however, that the discussion of dilaton stabilization 
in \cite{BW} which takes into accounts various observations of 
\cite{BaDi} as well as explicit mechanisms of dilaton stabilization \cite{kras}
suggests that to achieve the latter we do not seem to be allowed 
to have more than two large dimensions in the transverse directions. 
The recent discussion in \cite{AB} also suggests that radius stabilization 
at large values in the brane world context seems to favor having only two 
large directions. At any rate, at present it is not completely clear what 
the final resolution of all of these issues will be, so these questions still 
remain open.  

\subsection{``Bulk'' Gauge Symmetries}

{}In section IV we discussed various possibilities of having ``bulk'' gauge symmetries such as $U(1)_{\cal L}$ and $U(1)_{\cal R}$. We would like to finish this section by briefly pointing out how such gauge symmetries can arise in the brane world picture. 

{}At first it might seem that such gauge symmetries are not possible in, say, Type I context. Thus, the closed string sector, at least perturbatively, does not contain any $U(1)$ gauge fields: upon compactifying Type I on a Calabi-Yau three-fold (with $SU(3)$ holonomy) with Hodge numbers $(h^{1,1},h^{2,1})$ we have $h^{1,1}+h^{2,1}$ chiral multiplets and no vector multiplets in the closed string sector \cite{KST}. First, it is not completely clear whether this also holds non-perturbatively\footnote{Actually, from the dual heterotic picture we do expect that Abelian as well as non-Abelian ``bulk'' gauge symmetries could arise (non-perturbatively from the Type I viewpoint) at special points in the moduli space. These points correspond to the compactification sizes (or, more precisely, the K{\"a}hler structure) of order one. In the TeV-scale brane world scenarios we generically expect to be away from such special points which would imply that the corresponding continuous gauge symmetries are broken to their discrete subgroups.}. Second, the ``bulk'' gauge symmetries need {\em not} come from the closed string sector at all but could arise in the open string sector in the world-volumes of some higher dimensional branes (such as D9-branes). In fact, this way we could have not only Abelian but also non-Abelian ``bulk'' gauge symmetries. The latter might be important for understanding possible ways of suppressing FCNCs in the 
TeV-scale brane world scenarios \cite{flavor1,flavor} as well as for obtaining viable models of flavor hierarchy in this context.  

\section{Summary and Open Questions}

{}Let us briefly summarize the discussions in the previous sections. We have considered the TeV-scale Supersymmetric Standard Model proposed in \cite{zura}. The gauge coupling unification in the TSSM is as precise (at one loop) as in the MSSM, and occurs in the TeV range. The new states in this model which are crucial for the successful unification prediction turn out to be central in gauging anomaly free discrete (as well as continuous) symmetries which  forbid {\em all} of the dangerous higher dimensional baryon and lepton number violating operators and stabilize proton. At the same time these gauge symmetries protect the Majorana neutrino masses from being unacceptably large. 

{}The model we have considered in this paper, therefore, passes various non-trivial checks, and could be used as a testing ground for the TeV-scale brane world scenario. In the best case scenario we might even hope to be able to gradually build on this model toward a fully working model for the TeV-scale brane world scenario. One of the important open questions that needs to be addressed appears to be the issue of adequately suppressing flavor changing neutral currents \cite{flavor}. It would be interesting to see if one could incorporate a conclusive mechanism for suppressing FCNCs in the TSSM consistently with all the other features we have discussed in this paper.

{}In our discussion of possible brane world embeddings of the TSSM we have focused on Type I compactifications. It would be interesting to understand if one could obtain similar features in the heterotic M-theory framework \cite{HW} with the Standard Model gauge bosons localized on M-theory five-branes. The recent progress \cite{ovrut} in understanding and formulating systematic rules for constructing ${\cal N}=1$ supersymmetric heterotic M-theory vacua in four dimensions could facilitate model building in these directions. 

\acknowledgments

{}I would like to thank Nima Arkani-Hamed, Haim Goldberg, 
Pran Nath, Tom Taylor, Henry Tye, and especially Gia Dvali for 
valuable discussions. This work was supported in part by the grant 
NSF PHY-96-02074, 
and the DOE 1994 OJI award. I would also like to thank Albert and 
Ribena Yu for 
financial support.

\appendix
\section{Discrete Symmetries}

{}In this section we would like to review some facts about ``generation-blind'' discrete symmetries which we have used extensively in the main text of this paper. Here we will closely follow the corresponding discussion in \cite{IR}.

{}Thus, consider a ${\bf Z}_N$ symmetry acting on the chiral superfields $\phi_k$, $k=Q,D,U,L,E$. Let $g$ be the generator of ${\bf Z}_N$. The action of $g$ on $\phi_k$ 
is given by 
\begin{equation}
 g \phi_k =\exp(2\pi i \alpha_k/N)\phi_k~,
\end{equation}
where the ${\bf Z}_N$ charges $\alpha_k$ are integers. Once these five integers are specified, the corresponding ${\bf Z}_N$ charges $\alpha_{H_+}$ and $\alpha_{H_-}$  for the electroweak Higgs doublets follow from the requirement that the usual Yukawa couplings of quarks and leptons to $H_\pm$ be allowed. That is, we must have the couplings $QUH_+$, $QDH_-$ and $LEH_-$. The ${\bf Z}_N$ charge for the right handed neutrinos can also be determined by requiring that the usual coupling $LNH_+$ (which is important for neutrino mass generation) also be allowed. Note that the above five charges are not completely independent. They must satisfy the following constraint:
\begin{equation}    
 \alpha_Q+\alpha_D=\alpha_L+\alpha_E~.
\end{equation}
Finally, all ${\bf Z}_N$ elements which differ only by the discrete hypercharge rotations of the form $\exp[2\pi i (3Y)/N]$ are equivalent. This implies that shifting the charges $\alpha_k$ by
\begin{equation}\label{shift}    
 (\alpha_k)\rightarrow(\alpha_k)+(+1,+2,-4,-3,+6)
\end{equation}
can be used to set one of the charges, which we choose to be $\alpha_Q$, to zero.

{}It is then clear that there are only three independent ${\bf Z}_N$ symmetries we must consider. Let the corresponding generators be ${\cal L}$, ${\cal R}$ and ${\cal A}$. The corresponding charge assignments can conveniently be chosen as follows (the $({\cal L},{\cal R},{\cal A})$ charges are given in parentheses): 
\begin{eqnarray}
 &&Q:~(0,0,0)~,~~~D:~(0,+1,-1)~,~~~U:~(0,-1,0)~,\nonumber\\
 &&L:~(+1,0,-1)~,~~~E:~(-1,+1,0)~,~~~N:~(-1,-1,+1)~,\nonumber\\
 &&H_+:~(0,+1,0)~,~~~H_-:~(0,-1,+1)~.
\end{eqnarray} 
A general ${\bf Z}_N$ element can be written as ${\cal L}^m {\cal R}^n {\cal A}^p$. Note that any element with $p\not=0$ would forbid the $\mu$-term for the $H_\pm$ states. We will therefore only consider the elements containing ${\cal L}$ and/or ${\cal R}$ generators but not the ${\cal A}$ generator.

\end{document}